%% file: main.tex
\documentclass[journal]{IEEEtran}

\PassOptionsToPackage{hyphens}{url}

\usepackage{amsmath,amssymb,amsfonts,amsthm,mathrsfs}
\usepackage{adjustbox}
\usepackage{algorithm}
\usepackage{algpseudocode}
\usepackage{array}
\usepackage{booktabs}
\usepackage{cite}
\usepackage{graphicx}
\usepackage{stfloats}
\usepackage{tabularx}
\usepackage{textcomp}
\usepackage{xcolor}
\usepackage[hidelinks]{hyperref}

\hypersetup{breaklinks=true}

\input{Congmacros}

\newlength{\eqtagspace}
\newcommand{\eqfit}[1]{%
\adjustbox{max width=\dimexpr\linewidth-\eqtagspace\relax}{\(\displaystyle #1\)}%
}
\newcommand{\eqfitu}[1]{%
\adjustbox{max width=\linewidth}{\(\displaystyle #1\)}%
}

\newtheorem{proposition}{Proposition}
\newtheorem{corollary}{Corollary}
\newtheorem{lemma}{Lemma}
\newtheorem{definition}{Definition}
\newtheorem{remark}{Remark}

\newtheorem{assumption}{Assumption}

\newcommand*{\QED}{\hfill\ensuremath{\square}}

\begin{document}

\title{Mean-Field Learning for Storage Aggregation}

\author{Jingguan Liu,~Cong Chen,~Xiaomeng Ai,~Jiakun Fang,~Jinsong Wang,~and Jinyu Wen
	\thanks{This work was supported by the National Natural Science Foundation of China under Grant U25B6017.~\textit{(Corresponding authors:~Cong Chen and Xiaomeng Ai.)}}%
	\thanks{Jingguan Liu, Xiaomeng Ai, Jiakun Fang, and Jinyu Wen are with the State Key Laboratory of Advanced Electromagnetic Technology, Huazhong University of Science and Technology, Wuhan 430074, China (e-mail: \url{spencerplusmail@foxmail.com}; 
		\url{xiaomengai@hust.edu.cn}; \url{jfa@hust.edu.cn}; \url{jinyu.wen@hust.edu.cn}).}%
	\thanks{Cong Chen is with the Thayer School of Engineering, Dartmouth College, Hanover, NH 03755, USA (e-mail: \url{Cong.Chen@dartmouth.edu}).}%
	\thanks{Jinsong Wang is with HyperStrong Technology Co., Ltd., Beijing 100094, China (e-mail: \url{wangjinsong@hyperstrong.com}).}
}

\markboth{under review}%
{Liu \MakeLowercase{\textit{et al.}}: Mean-Field Learning for Storage Aggregation}

\maketitle

\begin{abstract}
Distributed energy storage devices can be aggregated to provide operational flexibility for power systems. Their efficient integration into system-level optimization calls for an accurate and computationally tractable surrogate that jointly represents aggregate flexibility and operating costs. However, constructing such a surrogate is challenging due to device heterogeneity, nonconvex operating constraints, and high-dimensional power trajectories. To address these challenges, this paper develops a mean-field learning framework for storage aggregation. Using random-set expectations, we show that the joint representation of aggregate flexibility and operating costs converges almost surely to a deterministic convex mean-field limit. This limit provides a learning benchmark for storage aggregation, against which we bound the approximation error of any convex surrogate.
These theoretical results motivate a convex virtual-battery surrogate with physically interpretable parameters. We learn these parameters through a bilevel optimization problem that matches the optimal values of the surrogate and the storage fleet when both respond to a set of price scenarios. We derive closed-form gradients to solve this problem efficiently. Case studies illustrate the theoretical results and demonstrate the framework's approximation accuracy, computational efficiency, and robustness to device heterogeneity.
\end{abstract}


\section{Introduction}\label{sec:intro}
\input{I_Intro}

\section{Problem Formulation}\label{sec:prob}
\input{II_Problem}

\section{Mean-Field Analysis of Storage Aggregation}\label{sec:MFanaly}

\input{III_MFanaly}

\section{Mean-Field Learning Framework}\label{sec:MFlearning}

\input{IV_MFlearn}

\section{Case Studies}\label{sec:case}
\input{V_Case}

\section{Conclusion}\label{sec:conclusion}
This paper proposes a mean-field learning framework for building a tractable surrogate model of a storage aggregator. Using random-set expectations, we show that the joint representation of aggregate flexibility and operating costs converges almost surely to a deterministic convex mean-field limit. We then bound the approximation error of any convex surrogate by its optimal-value gaps under a set of price scenarios plus a coverage term. Guided by these results, we learn a virtual-battery surrogate with physically interpretable parameters by matching its optimal values to those of the storage fleet, using closed-form gradients from the envelope theorem.

Case studies show that cooperation among storage devices makes the aggregate flexibility set nearly convex even at moderate scale, consistent with the mean-field analysis. The learned virtual battery generalizes to unseen prices with a median test error of 4.4\%, compared with 26--28\% for generic data-driven models. It is also robust to device heterogeneity: its median optimal-value gap stays below 1\% at all heterogeneity levels, whereas that of a model-based aggregation method grows to 20.1\% as the devices become more diverse.

\section*{Acknowledgments}
The authors would like to thank Prof. Eilyan Bitar from Cornell University for many insightful discussions.

\bibliographystyle{IEEEtran}
\bibliography{BIB}

\clearpage
\appendices

\input{Appendix}

\end{document}

%% file: Congmacros.tex
\def\beq{\begin{equation}}
\def\eeq{\end{equation}}
\def\bea{\begin{eqnarray}}
\def\eea{\end{eqnarray}}
\def\ba{\begin{array}}
\def\ea{\end{array}}

\def\bitem{\begin{itemize}}
\def\eitem{\end{itemize}}
\def\ben{\begin{enumerate}}
\def\een{\end{enumerate}}

\definecolor{bgrd}{rgb}{1,1,1}
\definecolor{gray}{rgb}{0.5,0.5,0.5}
\definecolor{dkr}{rgb}{0.7,0.1,0.2}
\definecolor{dkb}{rgb}{0.1,0.1,0.8}

\def\edoc{\end{document}}



%% file: I_Intro.tex
\IEEEPARstart{C}{oordinated} control of demand-side distributed energy resources can enhance power system flexibility \cite{wu2025optimal}. In particular, distributed energy storage devices---such as home backup batteries \cite{BasePower_Website}---can shift energy over time and support renewable integration. For example, Tesla's virtual power plant in California coordinates residential Powerwall batteries to support the grid during periods of system stress \cite{Tesla_VPP_PGE}. Policy initiatives such as FERC Order No.~2222 \cite{FERC_Order2222} further facilitate the participation of distributed energy resource aggregations in wholesale electricity markets \cite{chen2024wholesale}.
To participate in power system operations as a single virtual resource, a storage aggregator needs a tractable model suitable for system-level scheduling and dispatch \cite{CAISO_Sec30_2024}. This model should jointly characterize feasible aggregate power trajectories and the operating costs of realizing them, without explicitly representing individual devices in system-level optimization \cite{chen2026wholesale}.

Constructing such a model is challenging for large storage populations. Differences in device capacities, efficiencies, and initial energy states make aggregate behavior difficult to capture by simply scaling the model of a representative device. Discrete operating modes, including on/off decisions and mutually exclusive charging and discharging, introduce nonconvex constraints, further complicating the characterization of aggregate flexibility. Moreover, energy dynamics couple decisions across time, so exact aggregation must handle entire power trajectories, whose dimension grows with the scheduling horizon. These challenges motivate a compact, physically interpretable surrogate that accurately approximates aggregate flexibility and operating costs while remaining tractable for power system optimization.

\subsection{Related Work}
The flexibility of each storage device can be represented as a set of feasible power trajectories, with aggregate flexibility given by the Minkowski sum of the individual sets \cite{liu2025coupling}. Adding a cost dimension enables a joint representation of flexibility and operating costs, where each feasible aggregate trajectory is assigned the minimum combined operating cost over all feasible device-level schedules that realize it \cite{long2025characterizing}. Although exact characterizations of aggregate flexibility have been developed \cite{panda2024efficient,angeli2023exact}, they rely on specific structural assumptions, such as unidirectional operation, that limit their applicability to more general storage models. Explicit Minkowski-sum computation remains demanding for heterogeneous populations with nonconvex operating constraints and high-dimensional power trajectories \cite{liu2024continuous}.

Consequently, a variety of model-based approximation methods have been developed to obtain tractable surrogates. A common approach is to approximate each device's flexibility set as tightly as possible using a predefined geometric template, such as boxes \cite{chen2019aggregate}, ellipsoids \cite{chen2021leveraging}, zonotopes \cite{muller2017aggregation}, homothetic polytopes \cite{liu2025preference}, affine polytopes \cite{al2024efficient}, and vertex-hull representations \cite{ozturk2024alleviating}. These templates support efficient Minkowski-sum computation and thereby scale well to large populations.
Despite their computational appeal, two limitations remain: (i) device-level approximation errors can limit aggregate accuracy, especially in highly heterogeneous populations, and (ii) efficient approximation typically requires each device's flexibility set to be convex, an assumption violated by discrete on/off decisions and mutually exclusive charging and discharging.

Beyond these model-based methods, recent studies have explored data-driven aggregation using classifiers and neural networks \cite{lin2025model,he2026analyzing,hu2026constraint,taheri2022data}. These methods directly learn representations of aggregate feasible regions from operating data, often using feasibility labels or margins associated with sampled operating points. However, they provide limited theoretical insight into the joint structure of aggregate flexibility and operating costs as the storage population grows. Moreover, their benefits for storage aggregation may be limited because (i) learning detailed feasible boundaries can require substantial labeled data, particularly as the trajectory dimension increases, and (ii) neural-network parameters generally lack direct physical meaning, complicating model interpretation and validation for power system operations.

To address these limitations, we develop a mean-field learning framework that offers both theoretical insight and physical interpretability. We represent a large population of storage devices by a single \emph{mean-field aggregate} that summarizes their empirical average behavior and characterize its \emph{mean-field limit} as the population grows. This limit reveals the joint structure of aggregate flexibility and operating costs and guides surrogate learning. Mean-field ideas have been explored in variational inference \cite{han2019statistical}, game-theoretic equilibrium analysis \cite{guo2019learning}, and pricing for home energy management \cite{mehrabi2024optimal}. Our random-set formulation of storage aggregation connects population-level structure with physically interpretable surrogate learning, a connection that has received limited attention in prior work.

\subsection{Main Contributions}
This paper proposes a mean-field learning framework to develop \emph{an accurate and computationally efficient aggregation model} for large storage populations with heterogeneous parameters, nonconvex operating constraints, and high-dimensional power trajectories. The main contributions are threefold:

\subsubsection{Theoretical Analysis}
To our knowledge, this is the first work to develop a mean-field learning framework for large-scale storage aggregation based on random-set expectations. Under an i.i.d.\ population model, we show in Proposition~\ref{prop1:MFConverge} that the joint representation of aggregate flexibility and operating costs converges almost surely to a deterministic convex mean-field limit.  This limit provides a theoretical basis for representing the storage aggregator by a single convex population-level model, even though individual storage devices have heterogeneous, nonconvex operating constraints. Using this convexity, we derive in Proposition~\ref{prop2:DistanceBound} an error bound for any convex surrogate of the limit, which motivates learning the surrogate from the price responses of the storage fleet.

\subsubsection{Aggregation Method}
Building on the theoretical results, we design a convex virtual-battery surrogate for the storage aggregator with physically interpretable parameters. Guided by the error bound, we learn these parameters by matching the optimal values of the surrogate to those of the storage fleet when both respond to a set of price scenarios. The fleet's optimal values are obtained exactly by averaging device-level optimal values (Proposition~\ref{prop3:MFConsistence}), without explicitly computing high-dimensional Minkowski sums. We derive closed-form gradients via the envelope theorem, so each iteration only solves small quadratic programs (QPs) in parallel.

\subsubsection{Simulation Findings}
Our case studies yield three insights. First, cooperation among storage devices compensates for their individual nonconvexities. No single storage device can charge and discharge at once, but a fleet can mix idle, charging, and discharging devices, so the aggregate flexibility set becomes nearly convex. Second, a single virtual battery with a few physical parameters suffices to represent the fleet. It is thus learned efficiently from limited price data and generalizes to unseen prices far better than generic data-driven models, with a median test error of 4.4\% versus 26--28\%. Third, learning the surrogate from the fleet as a whole makes it robust to device heterogeneity. Its median optimal-value gap stays below 1\% at all heterogeneity levels, whereas that of a model-based method that approximates each device separately grows to 20.1\% as the devices become more diverse.

\subsection{Paper Organization and Notation}
The paper is organized as follows. Section~\ref{sec:prob} presents the problem formulation. Section~\ref{sec:MFanaly} analyzes storage aggregation from a mean-field perspective. Section~\ref{sec:MFlearning} introduces a mean-field learning framework. Section~\ref{sec:case} presents case studies, and Section~\ref{sec:conclusion} concludes. 

Regarding notation, for any vector $\boldsymbol{x}\in\mathbb{R}^{n}$, $\|\boldsymbol{x}\|$ denotes the $\ell_{2}$ norm. For any nonempty compact set $\mathcal{X}\subset\mathbb{R}^n$, we define
$
\|\mathcal{X}\|\;:=\;\max_{\boldsymbol{x}\in \mathcal{X}}\|\boldsymbol{x}\|.
$
We use $\operatorname{Conv}(\cdot)$ and $\mathbb{E}[\cdot]$ to denote the convex hull and the expectation operator, respectively.
For sets $\{\mathcal{X}_i\subset\mathbb{R}^{n}\}_{i=1}^{I}$, their Minkowski sum is defined as
\[
\bigoplus_{i=1}^{I}\mathcal{X}_i
:=\left\{\boldsymbol{x}\,\middle|\,
\boldsymbol{x}=\sum_{i=1}^{I}\boldsymbol{x}_i,\ \boldsymbol{x}_i\in\mathcal{X}_i\right\}\subset\mathbb{R}^{n}.
\]
Functions may take the value $+\infty$, and a minimum over an empty set is $+\infty$. The domain of $f:\mathbb{R}^n\to\mathbb{R}\cup\{+\infty\}$ is $\operatorname{dom}f:=\{\boldsymbol x\mid f(\boldsymbol x)<+\infty\}$. Given a common finite level $\bar u$, the uniformly truncated epigraph of $f$ is
\[
\operatorname{epi}_{\bar u} f
:=\left\{(\boldsymbol{x},u)\in\mathbb{R}^{n+1}\,\middle|\,
f(\boldsymbol{x})\le u\le\bar u\right\}.
\]
The same value of $\bar u$ is used for all device, aggregate, and surrogate models.
The closed Euclidean ball of radius $R\ge 0$ centered at $\boldsymbol{c}\in\mathbb{R}^{n}$ is
$
\mathcal{B}(\boldsymbol{c},R):=\left\{\boldsymbol{x}\,\middle|\,
\|\boldsymbol{x}-\boldsymbol{c}\|\le R\right\}\subset\mathbb{R}^{n}.
$
For nonempty compact sets $\mathcal{X}_1,\mathcal{X}_2\subset\mathbb{R}^{n}$, the Hausdorff distance is
\[
\eqfitu{
	d_{\mathrm{H}}(\mathcal{X}_1,\mathcal{X}_2)
	:=\min\left\{\varepsilon\ge 0\,\middle|\,
	\mathcal{X}_1\subseteq \mathcal{X}_2\mathbin{\oplus}\mathcal{B}(\boldsymbol 0,\varepsilon),\
	\mathcal{X}_2\subseteq \mathcal{X}_1\mathbin{\oplus}\mathcal{B}(\boldsymbol 0,\varepsilon)\right\}.
}
\]

%% file: II_Problem.tex
This section formulates the device and aggregate models and outlines how the subsequent sections construct a tractable convex surrogate for the storage aggregator.

\subsection{Individual Device Model}\label{subsec:problem_individual}
Consider a fleet of storage devices, indexed by $i\in \mathcal I:=\left\{1,\ldots,I\right\}$, over discrete time periods $t\in \mathcal T:=\left\{1,\ldots,T\right\}$ with period length $\Delta$. For all $t\in\mathcal T$, the operating constraints of each device are given by
\begin{subequations} \label{eq:ESS_ind}
	\begin{gather}
		e_{i,t}=e_{i,t-1}+\Delta\left(p_{i,t}^\mathrm C\eta_i^\mathrm C-p_{i,t}^\mathrm D/\eta_i^\mathrm D\right), \label{eq:ESS_ind_a} \\
		\underline{e}_{i}\le e_{i,t}\le\overline{e}_{i},~e_{i,0}=e_i^{\mathrm{ini}}, \label{eq:ESS_ind_b} \\
		\underline{p}_{i}\delta_{i,t}^\mathrm{C}\le p_{i,t}^\mathrm{C}\le \overline{p}_{i}\delta_{i,t}^\mathrm{C},\quad
		\underline{p}_{i}\delta_{i,t}^\mathrm{D}\le p_{i,t}^\mathrm{D}\le \overline{p}_{i}\delta_{i,t}^\mathrm{D},\label{eq:ESS_ind_c}\\
		\delta_{i,t}^\mathrm{C},\delta_{i,t}^\mathrm{D}\in \{0,1\},\quad
		\delta_{i,t}^\mathrm{C}+\delta_{i,t}^\mathrm{D}\le 1.\label{eq:ESS_ind_d}
	\end{gather}
\end{subequations}
Here, the decision variables are the charging power $p_{i,t}^{\mathrm C}$, discharging power $p_{i,t}^{\mathrm D}$, stored energy $e_{i,t}$, and binary on/off states $\delta_{i,t}^{\mathrm C}$ and $\delta_{i,t}^{\mathrm D}$. Parameters $\eta_i^\mathrm{C},\eta_i^\mathrm{D}\in(0,1]$ are the charging and discharging efficiencies, $\underline{e}_i,\overline{e}_i$ are energy capacity bounds, $e_i^{\mathrm{ini}}$ is the initial stored energy, and $\underline{p}_i,\overline{p}_i$ are the minimum and maximum charging or discharging power.

Let $\boldsymbol{p}_{i}^{\mathrm{C}}:=\left[ p_{i,t}^{\mathrm{C}} \right] _{t\in \mathcal T}$ and $\boldsymbol{p}_{i}^{\mathrm{D}}:=\left[ p_{i,t}^{\mathrm{D}} \right] _{t\in \mathcal T}$, and define the net power by $\boldsymbol p_i:=\boldsymbol p_i^{\mathrm C}-\boldsymbol p_i^{\mathrm D}\in\mathbb R^T$, where positive net power denotes charging. Let $\boldsymbol y_i:=[(\boldsymbol p_i^{\mathrm C})^\top \ (\boldsymbol p_i^{\mathrm D})^\top]^\top\in\mathbb R^{2T}$ be the stacked charge--discharge trajectory, and denote by $\mathcal Y_i^{\mathrm E}$ the set of $\boldsymbol y_i$ for which there exist $\boldsymbol e_i$, $\boldsymbol\delta_i^{\mathrm C}$, and $\boldsymbol\delta_i^{\mathrm D}$ satisfying \eqref{eq:ESS_ind}. The individual flexibility set in net-power coordinates is then
\beq \label{eq:ESS_ind_comp}
\mathcal{P} _{i}^{\mathrm{E}}:=\left\{ \boldsymbol{p}_i\in \mathbb{R} ^T\, \middle| \,\exists \boldsymbol{y}_i\in \mathcal{Y} _{i}^{\mathrm{E}}:
\boldsymbol{p}_{i}^{\mathrm{C}}-\boldsymbol{p}_{i}^{\mathrm{D}}=\boldsymbol{p}_i \right\} .
\eeq
The set $\mathcal P_i^{\mathrm E}$ is compact and, since the idle schedule is feasible, nonempty. In general, however, it is nonconvex because the binary variables in \eqref{eq:ESS_ind_c}--\eqref{eq:ESS_ind_d} impose a minimum operating power and mutually exclusive charging and discharging.

The total cost of each device depends on both the external electricity price and internal factors such as battery degradation. For a price vector $\boldsymbol\lambda\in\mathbb R^T$, any feasible profile $\boldsymbol p_i\in\mathcal P_i^{\mathrm E}$ incurs the total cost
\begin{equation} \label{eq:ESS_ind_costfunc}
	C_i^{\mathrm E}(\boldsymbol p_i,\boldsymbol\lambda)
	:=\boldsymbol\lambda^\top\boldsymbol p_i+U_i^{\mathrm E}(\boldsymbol p_i),
\end{equation}
where the first term is the price-dependent energy payment and the second term is a price-independent disutility, i.e., the internal operating cost of realizing $\boldsymbol p_i$. Following the degradation cost in \cite{tan2017optimal}, we adopt a linear-quadratic cost of the discharging power:
\begin{equation} \label{eq:ESS_ind_utilfunc}
	\eqfitu{
		U_{i}^{\mathrm{E}}(\boldsymbol{p}_i):=\min_{\substack{\boldsymbol y_i\in\mathcal Y_i^{\mathrm E}\\
				\boldsymbol p_i^{\mathrm C}-\boldsymbol p_i^{\mathrm D}=\boldsymbol p_i}}
		\sum_{t\in\mathcal T}\bigl(q_i^{\mathrm D}p_{i,t}^{\mathrm D}
		+Q_i^{\mathrm D}(p_{i,t}^{\mathrm D})^2\bigr),
	}
\end{equation}
where $q_i^{\mathrm D}\ge0$ and $Q_i^{\mathrm D}\ge0$ are the linear and quadratic discharge cost coefficients of device~$i$. By the convention on empty minima, $\operatorname{dom}U_i^{\mathrm E}=\mathcal P_i^{\mathrm E}$. We refer to the pair $(\mathcal P_i^{\mathrm E},U_i^{\mathrm E})$ as the \emph{model} of device~$i$.

\subsection{Aggregate Model}\label{subsec:problem_aggregate}
The \textit{aggregate flexibility set} represents all achievable average net-power trajectories:
\begin{equation}
	\eqfitu{
		\mathcal P_I^{\mathrm M}
		:=\frac1I\bigoplus_{i=1}^I\mathcal P_i^{\mathrm E}
		=\left\{\boldsymbol p\,\middle|\,
		\boldsymbol p=\frac1I\sum_{i=1}^I\boldsymbol p_i,\
		\boldsymbol p_i\in\mathcal P_i^{\mathrm E}\right\}\subset\mathbb R^T.
	}
	\label{eq:ESS_agg_set}
\end{equation}
The \textit{aggregate cost function} is
\begin{equation}
	\eqfitu{
		C_I^{\mathrm M}(\boldsymbol p,\boldsymbol\lambda)
		:=\boldsymbol\lambda^\top\boldsymbol p+U_I^{\mathrm M}(\boldsymbol p),
		\qquad \boldsymbol p\in\mathcal P_I^{\mathrm M},
	}
	\label{eq:ESS_agg_cost}
\end{equation}
where $U_I^{\mathrm M}$ is the minimum average disutility, with $\operatorname{dom}U_I^{\mathrm M}=\mathcal P_I^{\mathrm M}$:
\begin{equation}\label{eq:ESS_agg_util}
	\eqfitu{
		U_I^{\mathrm M}(\boldsymbol p)
		:=\min_{\{\boldsymbol p_i\}_{i=1}^I}\left\{
		\frac1I\sum_{i=1}^I U_i^{\mathrm E}(\boldsymbol p_i)
		\,\middle|\,
		\boldsymbol p=\frac1I\sum_{i=1}^I\boldsymbol p_i,\
		\boldsymbol p_i\in\mathcal P_i^{\mathrm E}\right\}.
	}
\end{equation}
We refer to $(\mathcal P_I^{\mathrm M},U_I^{\mathrm M})$ as the finite-population \emph{mean-field aggregate}, which describes per-device averages. Direct computation of \eqref{eq:ESS_agg_set}--\eqref{eq:ESS_agg_util} is intractable for large heterogeneous fleets. Our objective is thus to obtain a tractable convex surrogate approximation $(\widehat{\mathcal P}_\Theta,\widehat U_\Theta)$ of this pair.

When devices are modeled as independent random draws from a common distribution (Assumption~\ref{assumpt1:IndSetCost}), the mean-field aggregate $(\mathcal P_I^{\mathrm M},U_I^{\mathrm M})$ converges almost surely, as $I\to\infty$, to a deterministic convex limit $(\mathcal P^{\mathrm L},U^{\mathrm L})$, called the \emph{mean-field limit}. For large $I$, approximating the mean-field aggregate thus amounts to approximating this limit.

\subsection{Roadmap and Theoretical Motivation}\label{subsec:problem_roadmap}
\begin{figure}[!t]
	\centering
	\includegraphics[width=\linewidth]{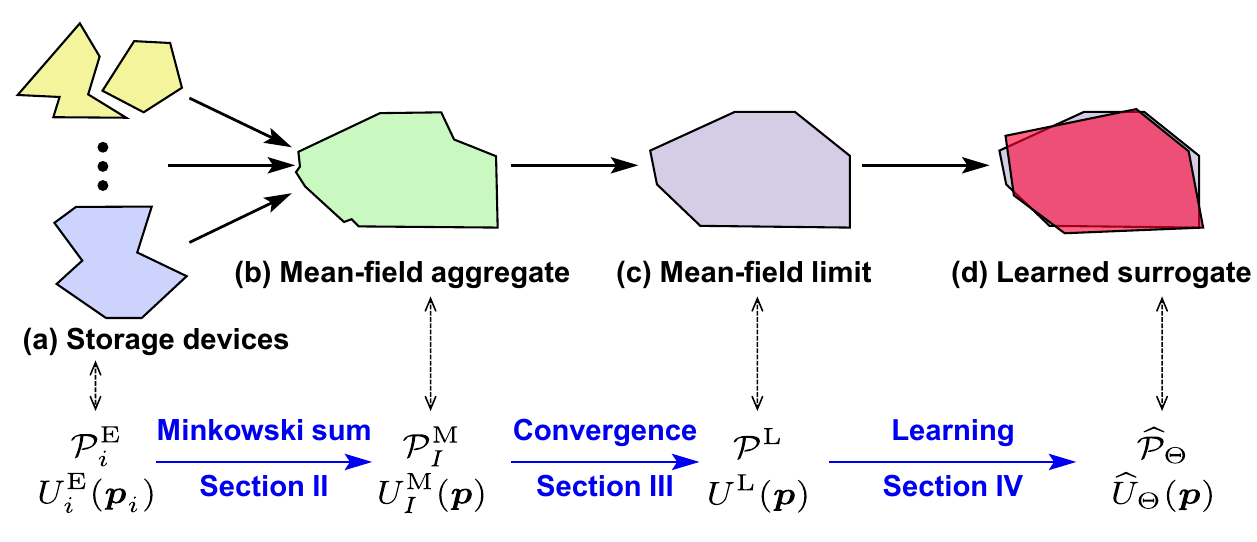}
	\caption{Relations among (a) the individual storage device models, (b) the finite-population mean-field aggregate of the storage fleet, (c) the mean-field limit, and (d) the surrogate learned from finite-population data to approximate this limit, with the sections that develop each step.}
	\label{fig:MethodOverview}
\end{figure}

Fig.~\ref{fig:MethodOverview} illustrates the key objects: (a) the individual models $(\mathcal P_i^{\mathrm E},U_i^{\mathrm E})$, (b) the mean-field aggregate $(\mathcal P_I^{\mathrm M},U_I^{\mathrm M})$, (c) the mean-field limit $(\mathcal P^{\mathrm L},U^{\mathrm L})$, and (d) the learned surrogate $(\widehat{\mathcal P}_\Theta,\widehat U_\Theta)$. Section~\ref{sec:MFanaly} first shows that the mean-field limit is convex (Proposition~\ref{prop1:MFConverge}), which motivates the convex surrogate design in this paper. We then compare the surrogate with the limit through their optimal values, i.e., their minimum total costs at given prices. The surrogate's approximation error is bounded by the optimal-value gaps under a set of price scenarios plus a coverage term (Proposition~\ref{prop2:DistanceBound}), and the optimal values of the limit are consistently estimated by averaging device-level optimal values (Proposition~\ref{prop3:MFConsistence}). Section~\ref{sec:MFlearning} designs the loss function of mean-field learning by matching the optimal values of the surrogate to these averages.

%% file: III_MFanaly.tex
This section conducts a theoretical analysis of storage aggregation to support the learning framework in Section~\ref{sec:MFlearning}.

\subsection{Preliminaries on Random Sets}\label{subsec:MFperformance_Preliminaries}
To analyze the aggregate model, we represent the individual storage device models by random sets, so that finite average Minkowski sums are empirical means of random sets and, by a law of large numbers, converge to a random-set expectation.

We first recall two definitions and a classical lemma. For an intuitive understanding of random sets, Appendix~\ref{subsec:appendix_comparison} provides a brief comparison between random variables and random sets. For a comprehensive treatment of random-set theory, we refer readers to \cite{molchanov2005theory, aumann1965integrals, artstein1975strong}.

\begin{definition}[Random Set \cite{molchanov2005theory}]\label{def1:RandomSet}
	A random compact set $\mathcal X\subset\mathbb R^n$ is a measurable random object whose realizations are nonempty compact subsets of $\mathbb R^n$.
\end{definition}

\begin{definition}[Aumann Expectation \cite{aumann1965integrals}]\label{def2:SetExpection}
	For an integrably bounded random compact set $\mathcal X$, i.e., $\mathbb E[\|\mathcal X\|]<\infty$,
	\[
	\begin{aligned}
		\mathbb E[\mathcal X]:=\{\,\mathbb E[\boldsymbol x]\mid {}&
		\boldsymbol x\text{ is an integrable}\\
		&\text{measurable selection of }\mathcal X\,\}.
	\end{aligned}
	\]
\end{definition}

\begin{lemma}[Strong Law for Random Sets \cite{artstein1975strong}]\label{lemma1:LargeNumber}
	Let $\{\mathcal X_i\}_{i\ge1}$ be i.i.d.\ integrably bounded random compact sets. Then
	\[
	d_{\mathrm H}\!\left(\frac1I\bigoplus_{i=1}^I\mathcal X_i,
	\mathbb E[\operatorname{Conv}(\mathcal X_1)]\right)
	\xrightarrow[I\to\infty]{\mathrm{a.s.}}0.
	\]
\end{lemma}

Lemma~\ref{lemma1:LargeNumber} supplies a deterministic population limit: finite average Minkowski sums converge almost surely to the Aumann expectation of the convexified random set. This is the key tool of our mean-field view of storage aggregation.

\subsection{Existence and Convexity of Mean-Field Limits}\label{subsec:MFperformance_Existence}
We now apply Lemma~\ref{lemma1:LargeNumber} to the storage aggregation problem. We impose the following assumption directly on the device models rather than on their parameters.

\begin{assumption}[I.i.d. Device Models]\label{assumpt1:IndSetCost}
	We assume that the sequence of device models is defined on one common probability space and satisfies the following two conditions. (a)~\emph{Bounded disutility}: there is a deterministic common truncation level $\bar u$ such that, almost
	surely, for every $i\ge1$,
	\begin{equation}\label{eq:common_upper}
		\sup_{\boldsymbol p\in\mathcal P_i^{\mathrm E}}
		U_i^{\mathrm E}(\boldsymbol p)<\bar u.
	\end{equation}
	(b)~\emph{I.i.d.\ augmented sets}: the augmented sets $\{\mathcal E_i\}_{i\ge1}$, defined by
	\begin{equation}\label{eq:Epi_ESSind}
		\begin{aligned}
			\mathcal E_i&:=\operatorname{epi}_{\bar u}U_i^{\mathrm E}\\
			&=\{(\boldsymbol p,u)\in\mathbb R^{T+1}\mid
			\boldsymbol p\in\mathcal P_i^{\mathrm E},\
			U_i^{\mathrm E}(\boldsymbol p)\le u\le\bar u\},
		\end{aligned}
	\end{equation}
	are i.i.d.\ measurable random
	compact sets and are integrably bounded, i.e.,
	$\mathbb E[\|\mathcal E_1\|]<\infty$.
\end{assumption}
The augmented set $\mathcal E_i$ in~\eqref{eq:Epi_ESSind} jointly represents the flexibility and disutility of device~$i$. It pairs every feasible power profile $\boldsymbol p$ with every cost level $u$ between its disutility $U_i^{\mathrm E}(\boldsymbol p)$ and $\bar u$. Truncation at $\bar u$ keeps $\mathcal E_i$ compact, as Lemma~\ref{lemma1:LargeNumber} requires. Condition~\eqref{eq:common_upper} ensures that this truncation excludes no feasible profile. This condition is reasonable because the power limits and cost coefficients of storage devices are bounded, and so are their disutilities.

\begin{remark}[Scope of Assumption~\ref{assumpt1:IndSetCost}]\label{rem:AssumptionScope}
For the storage device model \eqref{eq:ESS_ind}, Assumption~\ref{assumpt1:IndSetCost} holds whenever the parameter vectors of the devices are i.i.d.\ and uniformly bounded (Appendix~\ref{subsec:appendix_params}). This setting allows heterogeneous storage fleets: capacities, power limits, and efficiencies differ across devices, and the parameters of the same device may be correlated with each other. It also covers fleets with several product types, since the type of a device can be treated as one more random parameter. The assumption is needed only for the deterministic mean-field limit in Propositions~\ref{prop1:MFConverge}--\ref{prop3:MFConsistence}.
It fails if groups of devices share random factors, as under strong spatial or demographic clustering, in which case the mean-field aggregate need not converge to a deterministic limit. The learning method in Section~\ref{sec:MFlearning}, however, does not require the i.i.d.\ condition of Assumption~\ref{assumpt1:IndSetCost} and retains an error bound for the fleet it is learned from (Remark~\ref{rem1:LearningScope}).
\end{remark}

Because the truncation level is common and exceeds every device disutility, the following identity holds almost surely for every finite $I$:
\begin{equation}\label{eq:epi_minkowski_identity}
	\operatorname{epi}_{\bar u}U_I^{\mathrm M}
	=\frac1I\bigoplus_{i=1}^I\operatorname{epi}_{\bar u}U_i^{\mathrm E}
	=\frac1I\bigoplus_{i=1}^I\mathcal E_i.
\end{equation}

For a set $\mathcal E\subset\mathbb R^{T+1}$ with coordinates $(\boldsymbol p,u)$, the projection that drops the cost coordinate $u$ is
\[
\operatorname{proj}_{\boldsymbol p}\mathcal E
:=\{\boldsymbol p\in\mathbb R^T\mid \exists u\in\mathbb R:
(\boldsymbol p,u)\in\mathcal E\}.
\]
Set
\[
\mathcal E^{\mathrm L}:=\mathbb E[\operatorname{Conv}(\mathcal E_1)],
\qquad
\mathcal P^{\mathrm L}:=\operatorname{proj}_{\boldsymbol p}\mathcal E^{\mathrm L}.
\]

\begin{proposition}[Mean-Field Limit of Storage Aggregation]\label{prop1:MFConverge}
	Under Assumption~\ref{assumpt1:IndSetCost}, $\mathcal E^{\mathrm L}$ and $\mathcal P^{\mathrm L}$ are nonempty compact convex sets. There is a unique lower-semicontinuous convex function $U^{\mathrm L}$ with $\operatorname{dom}U^{\mathrm L}=\mathcal P^{\mathrm L}$ such that
	\begin{equation}\label{eq:prop1_epi}
		\mathcal E^{\mathrm L}=\operatorname{epi}_{\bar u}U^{\mathrm L},
		\qquad
		d_{\mathrm H}\!\left(\operatorname{epi}_{\bar u}U_I^{\mathrm M},
		\mathcal E^{\mathrm L}\right)
		\xrightarrow[I\to\infty]{\mathrm{a.s.}}0.
	\end{equation}
	Consequently, projection gives
	\begin{equation}\label{eq:prop1_set}
		d_{\mathrm H}(\mathcal P_I^{\mathrm M},\mathcal P^{\mathrm L})
		\xrightarrow[I\to\infty]{\mathrm{a.s.}}0,
	\end{equation}
	and the disutilities epi-converge:
	\begin{equation}\label{eq:prop1_epiconv}
		U_I^{\mathrm M}
		\xrightarrow[I\to\infty]{\mathrm{epi},\ \mathrm{a.s.}}
		U^{\mathrm L}.
	\end{equation}
\end{proposition}

The proof, given in Appendix~\ref{subsec:appendix_prop1}, applies Lemma~\ref{lemma1:LargeNumber} to the augmented sets $\mathcal E_i$ in~\eqref{eq:Epi_ESSind}.

Proposition~\ref{prop1:MFConverge} yields two insights for storage aggregation. First, the mean-field limit $\mathcal E^{\mathrm L}$ is deterministic. It depends only on the common distribution of the storage devices, and so do the flexibility set $\mathcal P^{\mathrm L}$ and disutility $U^{\mathrm L}$ obtained from it. For almost every realization of the fleet, the mean-field aggregate converges to this same limit, so a single population-level model can represent a heterogeneous storage fleet. Second, the mean-field limit is convex even when individual flexibility sets and disutilities are nonconvex. Physically, a large fleet can mix idle, charging, and discharging devices, so its average power can take values that no single device can realize (Section~\ref{subsec:toyexample}). 

Together, these two insights reduce the aggregation of a large heterogeneous storage fleet to learning a single convex surrogate.

\subsection{Approximation Error via Price Responses}\label{subsec:MFperformance_Behavior}
We consider a convex surrogate parameterized by a vector $\Theta$. Let $\widehat{\mathcal P}_\Theta\subset\mathbb R^T$
be a nonempty compact convex surrogate set and let
$\widehat U_\Theta$ be a lower-semicontinuous convex function with
$\operatorname{dom}\widehat U_\Theta=\widehat{\mathcal P}_\Theta$. Assume that
\[
\sup_{\boldsymbol p\in\widehat{\mathcal P}_\Theta}
\widehat U_\Theta(\boldsymbol p)<\bar u,
\]
where $\bar u$ is the common level fixed in Assumption~\ref{assumpt1:IndSetCost}. Define
\begin{equation}\label{eq:surrogate_epi_price}
\widehat{\mathcal E}_\Theta:=\operatorname{epi}_{\bar u}\widehat U_\Theta,
\qquad
\boldsymbol\pi(\boldsymbol\lambda):=[\boldsymbol\lambda^\top\ 1]^\top.
\end{equation}
For $\boldsymbol z=[\boldsymbol p^\top\ u]^\top\in\mathbb R^{T+1}$, the product $\boldsymbol\pi(\boldsymbol\lambda)^\top\boldsymbol z=\boldsymbol\lambda^\top\boldsymbol p+u$ is the total cost of the profile $\boldsymbol p$ at price $\boldsymbol\lambda$ with disutility $u$.

We measure the approximation accuracy of the convex storage surrogate $\widehat{\mathcal E}_\Theta$ with respect to the mean-field limit $\mathcal E^{\mathrm L}$ by the Hausdorff distance between them, which captures flexibility and disutility errors jointly. However, this distance cannot be evaluated directly because $\mathcal E^{\mathrm L}$ generally has no analytical expression. We therefore compare the two models through their responses to prices. Given a price $\boldsymbol\lambda$, each model solves a \emph{price-response problem} that minimizes its total cost, with the \emph{optimal values}
\begin{equation}\label{eq:support_functions}
	\begin{aligned}
		V^{\mathrm L}(\boldsymbol\lambda)
		&:=\min_{\boldsymbol z\in\mathcal E^{\mathrm L}}
		\boldsymbol\pi(\boldsymbol\lambda)^\top\boldsymbol z
		=\min_{\boldsymbol p\in\mathcal P^{\mathrm L}}
		\bigl\{\boldsymbol\lambda^\top\boldsymbol p+U^{\mathrm L}(\boldsymbol p)\bigr\},\\
		\widehat V_\Theta(\boldsymbol\lambda)
		&:=\min_{\boldsymbol z\in\widehat{\mathcal E}_\Theta}
		\boldsymbol\pi(\boldsymbol\lambda)^\top\boldsymbol z
		=\min_{\boldsymbol p\in\widehat{\mathcal P}_\Theta}
		\bigl\{\boldsymbol\lambda^\top\boldsymbol p+\widehat U_\Theta(\boldsymbol p)\bigr\}.
	\end{aligned}
\end{equation}
The second equalities hold because, by~\eqref{eq:surrogate_epi_price}, the last entry of $\boldsymbol\pi(\boldsymbol\lambda)$ is one, so each minimum is attained on the lower boundary of the truncated epigraph, where the cost level $u$ equals the disutility. In convex analysis, $V^{\mathrm L}$ and $\widehat V_\Theta$ are, up to a sign convention, the support functions of $\mathcal E^{\mathrm L}$ and $\widehat{\mathcal E}_\Theta$. Since both sets are convex, they are fully determined by their optimal values at all prices (Appendix~\ref{subsec:appendix_prop2}).

In practice, only finitely many prices can be examined. Consider a finite set of \emph{price scenarios} $\{\boldsymbol\lambda_d\}_{d=1}^D$, with $D\ge1$, and define the optimal-value gap at scenario $d$ as
\begin{equation}\label{eq:support_mismatch}
	\varepsilon_d:=\left|V^{\mathrm L}(\boldsymbol\lambda_d)
	-\widehat V_\Theta(\boldsymbol\lambda_d)\right|.
\end{equation}
Let $R<\infty$ bound the norms of all points $(\boldsymbol p,U^{\mathrm L}(\boldsymbol p))$, $\boldsymbol p\in\mathcal P^{\mathrm L}$, and $(\boldsymbol p,\widehat U_\Theta(\boldsymbol p))$, $\boldsymbol p\in\widehat{\mathcal P}_\Theta$, on the graphs of the two disutilities. Define the normalized price direction
\[
\boldsymbol v(\boldsymbol\lambda)
:=\frac{\boldsymbol\pi(\boldsymbol\lambda)}{\|\boldsymbol\pi(\boldsymbol\lambda)\|}=\frac{[\boldsymbol\lambda^\top\ 1]^\top}
{\sqrt{1+\|\boldsymbol\lambda\|^2}},
\]
and set
\[
\kappa_D:=
\sup_{\boldsymbol\lambda\in\mathbb R^T}
\min_{d=1,\ldots,D}
\|\boldsymbol v(\boldsymbol\lambda)-\boldsymbol v(\boldsymbol\lambda_d)\|.
\]
Thus, $\kappa_D$ measures how well the price scenarios cover all price directions. It is the largest distance from any price direction to the nearest scenario direction.

\begin{proposition}[Error Bound for Convex Surrogates]\label{prop2:DistanceBound}
	Under Assumption~\ref{assumpt1:IndSetCost}, let the surrogate satisfy the above conditions. Then, for any price scenarios $\{\boldsymbol\lambda_d\}_{d=1}^D$,
	\begin{equation}\label{eq:finite_epigraph_mse}
		d_{\mathrm H}(\mathcal E^{\mathrm L},\widehat{\mathcal E}_\Theta)
		\le\max_{d=1,\ldots,D}\varepsilon_d+2R\kappa_D.
	\end{equation}
\end{proposition}

The proof is given in Appendix~\ref{subsec:appendix_prop2}.

Proposition~\ref{prop2:DistanceBound} yields two insights for surrogate learning. First, the bound in \eqref{eq:finite_epigraph_mse} splits the approximation error into the optimal-value gaps at the price scenarios and a coverage term. The gaps are the part that learning can reduce, so Section~\ref{sec:MFlearning} learns the surrogate by minimizing them. Each optimal value requires only solving a price-response problem rather than characterizing the whole aggregate set. Second, since operating prices are bounded, the price scenarios cannot cover the directions of arbitrarily large prices, so the coverage term has a positive floor no matter how many scenarios are used. This floor reflects the worst case over all convex surrogates. In contrast, a surrogate with a physical structure, such as the virtual-battery model in Section~\ref{subsec:framework:surrogate}, is determined by only a few physical parameters, so a modest number of well-spread price scenarios typically suffices to learn it accurately, as illustrated in Section~\ref{sec:case}.

The gaps, however, involve the mean-field values $V^{\mathrm L}(\boldsymbol\lambda_d)$, which are not directly computable either. Section~\ref{subsec:MFperformance_response} replaces them by consistent finite-population reference values, and Section~\ref{sec:MFlearning} learns the surrogate from these values.

\subsection{Consistent Reference Optimal Values}\label{subsec:MFperformance_response}
For each price scenario $\boldsymbol\lambda_d$ in Section~\ref{subsec:MFperformance_Behavior}, we define in~\eqref{eq:ref_support_values} below a reference optimal value for the finite storage fleet. By~\eqref{eq:epi_minkowski_identity}, it is the optimal value of the price-response problem of the mean-field aggregate with the cost~\eqref{eq:ESS_agg_cost}. The following proposition shows how to compute it and establishes its consistency.

\begin{proposition}[Reference Optimal Values]\label{prop3:MFConsistence}
	Under Assumption~\ref{assumpt1:IndSetCost}, for every $I\ge1$ and every $d=1,\ldots,D$, the reference optimal value of the storage fleet satisfies, almost surely,
	\begin{equation}\label{eq:ref_support_values}
		\begin{aligned}
			V_{I,d}^{\mathrm{ref}}
			&:=\min_{\boldsymbol z\in\frac1I\bigoplus_{i=1}^I\mathcal E_i}
			\boldsymbol\pi(\boldsymbol\lambda_d)^\top\boldsymbol z\\
			&=\frac1I\sum_{i=1}^I\min_{\boldsymbol z\in\mathcal E_i}
			\boldsymbol\pi(\boldsymbol\lambda_d)^\top\boldsymbol z,
		\end{aligned}
	\end{equation}
	i.e., it is the average of the device-level optimal values. Moreover, for every $d$,
	\begin{equation}\label{eq:support_slln}
		V_{I,d}^{\mathrm{ref}}
		\xrightarrow[I\to\infty]{\mathrm{a.s.}}
		V^{\mathrm L}(\boldsymbol\lambda_d)
		=\mathbb E\Bigl[\min_{\boldsymbol z\in\mathcal E_1}
		\boldsymbol\pi(\boldsymbol\lambda_d)^\top\boldsymbol z\Bigr],
	\end{equation}
	where $V^{\mathrm L}(\boldsymbol\lambda_d)$ is the optimal value of the mean-field limit
	defined in~\eqref{eq:support_functions}.
\end{proposition}

The proof is given in Appendix~\ref{subsec:appendix_prop3}.

Proposition~\ref{prop3:MFConsistence} yields two insights for surrogate learning. First, the optimal value of the mean-field aggregate at a given price is the average of the device-level optimal values, because each device can respond to the price independently. The reference values are thus obtained exactly from only $I$ device-level price-response problems, without computing high-dimensional Minkowski sums. Second, the reference values converge to the mean-field values $V^{\mathrm L}(\boldsymbol\lambda_d)$ as the fleet grows. Learning the surrogate from the reference values of a large fleet therefore approximates learning it from the mean-field limit.

%% file: IV_MFlearn.tex
This section presents the learning framework, which consists of the surrogate design, the surrogate learning problem, and the solution algorithm.

\subsection{Framework Overview}
\begin{figure}[!t]
	\centering
	\includegraphics[width=\linewidth]{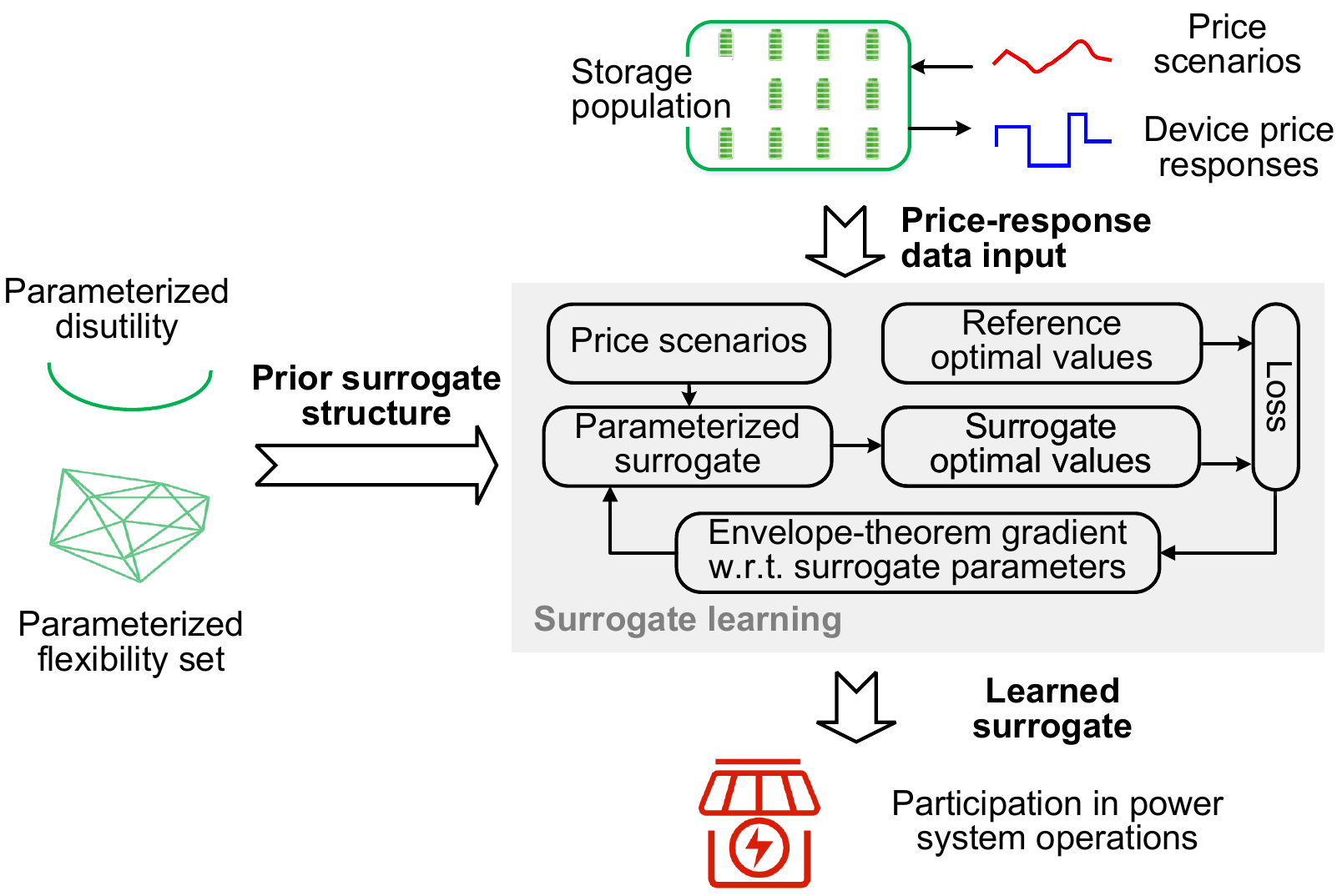}
	\caption{Flowchart of the proposed learning framework.}
	\label{fig:LearningFramework}
\end{figure}

Fig.~\ref{fig:LearningFramework} summarizes the proposed learning framework. Motivated by Proposition~\ref{prop1:MFConverge}, we adopt a convex virtual-battery surrogate that can be embedded directly in system-level optimization. It consists of a parameterized flexibility set and a parameterized disutility, shown in green in the left panel. At the top, the price responses of the storage devices to the price scenarios are computed centrally. Their averaged optimal values enter the gray-shaded surrogate-learning block as reference optimal values. By Proposition~\ref{prop3:MFConsistence}, these reference values are exact for the finite fleet and consistent with the mean-field limit. The block matches the optimal values of the surrogate to the reference values (Section~\ref{subsec:framework_problem}), which controls the Hausdorff distance to the mean-field limit (Corollary~\ref{cor1:LearningBound}). The gradient algorithm of Section~\ref{subsec:framework_solution} solves the surrogate learning problem.

\subsection{Population-Level Virtual-Battery Surrogate}\label{subsec:framework:surrogate}
We approximate the mean-field limit $(\mathcal P^{\mathrm L},U^{\mathrm L})$ by a convex surrogate with parameters
\[
\begin{aligned}
	\Theta:=\bigl(&\eta^{\mathrm C},\eta^{\mathrm D},
	\overline p^{\mathrm C},\overline p^{\mathrm D},e^{\mathrm{ini}},
	\underline e,\overline e,q^{\mathrm D},Q^{\mathrm D}\bigr).
\end{aligned}
\]
Here $\eta^{\mathrm C},\eta^{\mathrm D}\in(0,1]$ are the charging and discharging efficiencies, and $\overline p^{\mathrm C},\overline p^{\mathrm D}\ge0$ are the maximum charging and discharging powers. The stored energy starts at $e^{\mathrm{ini}}$ and stays between the bounds $\underline e$ and $\overline e$, where $\underline e\le e^{\mathrm{ini}}\le\overline e$. Finally, $q^{\mathrm D},Q^{\mathrm D}\ge0$ are the linear and quadratic discharge cost coefficients. The feasible charge--discharge trajectories $\boldsymbol y=[(\boldsymbol p^{\mathrm C})^\top\ (\boldsymbol p^{\mathrm D})^\top]^\top\in\mathbb R^{2T}$ of the surrogate form the convex set
\begin{equation}\label{eq:VB_internal}
	\begin{aligned}
		\mathcal Y_\Theta:=\{\,\boldsymbol y\mid {}&
		\exists(e_0,\ldots,e_T):\ e_0=e^{\mathrm{ini}},\\
		&e_t=e_{t-1}+\Delta\Bigl(\eta^{\mathrm C}p_t^{\mathrm C}
		-\frac{p_t^{\mathrm D}}{\eta^{\mathrm D}}\Bigr),\\
		&\underline e\le e_t\le\overline e,\quad
		0\le p_t^{\mathrm C}\le\overline p^{\mathrm C},\\
		&0\le p_t^{\mathrm D}\le\overline p^{\mathrm D},
		\quad t\in\mathcal T\,\}.
	\end{aligned}
\end{equation}
As in \eqref{eq:ESS_ind_comp} and \eqref{eq:ESS_ind_utilfunc}, the surrogate flexibility set is
\begin{equation}\label{eq:VB_set}
	\widehat{\mathcal P}_\Theta
	:=\{\boldsymbol p\in\mathbb R^T\mid
	\exists\boldsymbol y\in\mathcal Y_\Theta:\
	\boldsymbol p^{\mathrm C}-\boldsymbol p^{\mathrm D}=\boldsymbol p\},
\end{equation}
and the corresponding surrogate disutility, with $\operatorname{dom}\widehat U_\Theta=\widehat{\mathcal P}_\Theta$, is
\begin{equation}\label{eq:VB_cost}
	\eqfitu{
		\widehat U_\Theta(\boldsymbol p)
		:=\min_{\substack{\boldsymbol y\in\mathcal Y_\Theta\\
				\boldsymbol p^{\mathrm C}-\boldsymbol p^{\mathrm D}=\boldsymbol p}}
		\sum_{t\in\mathcal T}\bigl(q^{\mathrm D}p_t^{\mathrm D}
		+Q^{\mathrm D}(p_t^{\mathrm D})^2\bigr).
	}
\end{equation}

The surrogate is thus a single virtual battery whose parameters have the same physical meaning as those of a device in Section~\ref{subsec:problem_individual} but are learned from the per-device average behavior of the population. Unlike the device model \eqref{eq:ESS_ind}, it has no binary variables or minimum power limits, in line with the convexity of the mean-field limit established in Proposition~\ref{prop1:MFConverge} and illustrated in Section~\ref{subsec:toyexample}. Because its constraints are linear, its cost is convex, and the idle schedule $\boldsymbol y=\boldsymbol0$ is feasible, $\widehat{\mathcal P}_\Theta$ is a nonempty polytope and $\widehat U_\Theta$ is convex and lower semicontinuous. For $\bar u$ above the maximum of $\widehat U_\Theta$, the surrogate thus satisfies the conditions of Proposition~\ref{prop2:DistanceBound}.

\subsection{Surrogate Learning Problem}\label{subsec:framework_problem}
Motivated by Proposition~\ref{prop2:DistanceBound}, we learn $\Theta$ from price responses, by matching the optimal values of the surrogate to those of the storage fleet under the price scenarios $\{\boldsymbol\lambda_d\}_{d=1}^D$. For each device $i$ and price scenario $\boldsymbol\lambda_d$, let $\boldsymbol p_{i,d}^{\mathrm{ref}}$ be an optimal solution of the device-level price-response problem $\min_{\boldsymbol p_i\in\mathcal P_i^{\mathrm E}}C_i^{\mathrm E}(\boldsymbol p_i,\boldsymbol\lambda_d)$, a mixed-integer quadratic program (MIQP) over \eqref{eq:ESS_ind} with the cost \eqref{eq:ESS_ind_costfunc}. Its optimal value is the device-level minimum in \eqref{eq:ref_support_values}. By Proposition~\ref{prop3:MFConsistence}, the reference optimal value and the reference response of the fleet are the device averages
\begin{equation}\label{eq:ref_support_value}
	V_{I,d}^{\mathrm{ref}}
	=\frac1I\sum_{i=1}^I C_i^{\mathrm E}(\boldsymbol p_{i,d}^{\mathrm{ref}},\boldsymbol\lambda_d),
	\qquad
	\overline{\boldsymbol p}_{I,d}^{\mathrm{ref}}
	:=\frac1I\sum_{i=1}^I\boldsymbol p_{i,d}^{\mathrm{ref}}.
\end{equation}

For the same price scenario, the price-response problem of the surrogate is the QP
\begin{equation}\label{eq:solvesurrogate}
	\begin{aligned}
		\widehat V_\Theta(\boldsymbol\lambda_d)
		&=\min_{\boldsymbol y\in\mathcal Y_\Theta}\
		\boldsymbol\lambda_d^\top
		(\boldsymbol p^{\mathrm C}-\boldsymbol p^{\mathrm D})\\
		&\qquad{}+\sum_{t\in\mathcal T}\bigl(q^{\mathrm D}p_t^{\mathrm D}
		+Q^{\mathrm D}(p_t^{\mathrm D})^2\bigr),
	\end{aligned}
\end{equation}
which is equivalent to minimizing $\boldsymbol\lambda_d^\top\boldsymbol p+\widehat U_\Theta(\boldsymbol p)$ over $\widehat{\mathcal P}_\Theta$ and thus gives the surrogate optimal value in \eqref{eq:support_functions}. An optimal solution $\boldsymbol y_d^*(\Theta)=[(\boldsymbol p_d^{\mathrm C,*}(\Theta))^\top\ (\boldsymbol p_d^{\mathrm D,*}(\Theta))^\top]^\top$ yields the surrogate response $\widehat{\boldsymbol p}_d^*(\Theta):=\boldsymbol p_d^{\mathrm C,*}(\Theta)-\boldsymbol p_d^{\mathrm D,*}(\Theta)$, where the superscript $*$ marks optimal quantities.

The surrogate learning problem minimizes the mean squared optimal-value gap between the surrogate and the storage fleet:
\begin{equation}\label{eq:inverseoptimization}
	\begin{aligned}
		\min_{\Theta}\quad
		&\mathcal J(\Theta):=\frac1D\sum_{d=1}^D
		\left|V_{I,d}^{\mathrm{ref}}
		-\widehat V_\Theta(\boldsymbol\lambda_d)\right|^2\\
		\mathrm{s.t.}\quad
		&\text{\eqref{eq:solvesurrogate} for }d=1,\ldots,D.
	\end{aligned}
\end{equation}
The loss function $\mathcal J(\Theta)$ of mean-field learning controls the approximation error with respect to the mean-field limit as follows.
\begin{corollary}[Learning Error Bound]\label{cor1:LearningBound}
	Under Assumption~\ref{assumpt1:IndSetCost}, let the surrogate satisfy the conditions of Proposition~\ref{prop2:DistanceBound}. Then, for every $I\ge1$,
	\begin{equation}\label{eq:learning_bound}
		\eqfit{
			d_{\mathrm H}(\mathcal E^{\mathrm L},\widehat{\mathcal E}_\Theta)
			\le\sqrt{D\,\mathcal J(\Theta)}+2R\kappa_D
			+\max_{d}\bigl|V_{I,d}^{\mathrm{ref}}-V^{\mathrm L}(\boldsymbol\lambda_d)\bigr|,
		}
	\end{equation}
	where the last term converges to zero almost surely as $I\to\infty$.
\end{corollary}
The proof is given in Appendix~\ref{subsec:appendix_prop2}.

\begin{remark}[Scope of the Learning Method]\label{rem1:LearningScope}
The learning method does not require the i.i.d.\ condition of Assumption~\ref{assumpt1:IndSetCost}. Problem~\eqref{eq:inverseoptimization} uses only the data of the given storage fleet, and its accuracy guarantee requires only the common truncation level in~\eqref{eq:common_upper}. Under~\eqref{eq:common_upper}, the reference values $V_{I,d}^{\mathrm{ref}}$ are the exact optimal values of the fleet's own aggregate model $\operatorname{epi}_{\bar u}U_I^{\mathrm M}$ and, because the total cost is linear in $(\boldsymbol p,u)$, also of its convex hull $\mathcal H_I:=\operatorname{Conv}(\operatorname{epi}_{\bar u}U_I^{\mathrm M})$. The proof of Proposition~\ref{prop2:DistanceBound} uses only convexity and the common truncation level, so it applies with $\mathcal H_I$ in place of $\mathcal E^{\mathrm L}$ and gives $d_{\mathrm H}(\mathcal H_I,\widehat{\mathcal E}_\Theta)\le\sqrt{D\,\mathcal J(\Theta)}+2R\kappa_D$, i.e., \eqref{eq:learning_bound} without the finite-population term. By the Shapley--Folkman--Starr theorem~\cite{starr1969quasi}, $d_{\mathrm H}(\mathcal H_I,\operatorname{epi}_{\bar u}U_I^{\mathrm M})\le\sqrt{T+1}\max_i\|\mathcal E_i\|/I$, so the learned surrogate also approximates the aggregate model of a large fleet itself. Assumption~\ref{assumpt1:IndSetCost} adds generalization across fleets: by Proposition~\ref{prop1:MFConverge}, $\mathcal H_I$ converges to the mean-field limit $\mathcal E^{\mathrm L}$, so a single surrogate represents all large fleets drawn from the same distribution. Without this assumption, the surrogate represents only the fleet it is learned from. Appendix~\ref{subsec:appendix_noniid} confirms numerically that the learning method remains accurate for fleets of correlated devices.
\end{remark}

\subsection{Closed-Form Gradient and Parallel Solution}\label{subsec:framework_solution}
Problem~\eqref{eq:inverseoptimization} is a bilevel program, in which each $\widehat V_\Theta(\boldsymbol\lambda_d)$ is the optimal value of the lower-level QP~\eqref{eq:solvesurrogate}. Since the loss depends on the QPs only through these optimal values, we derive its gradient in closed form by the chain rule and the envelope theorem~\cite{bonnans2000perturbation}:
\begin{equation}\label{eq:gradient}
	\begin{aligned}
		\nabla_\Theta\mathcal J(\Theta)
		=-\frac2D\sum_{d=1}^D{}&\bigl(V_{I,d}^{\mathrm{ref}}-\widehat V_\Theta(\boldsymbol\lambda_d)\bigr)\\
		&\times\nabla_\Theta\mathcal L_d(\boldsymbol x_d^*,\boldsymbol\nu_d^*,\boldsymbol\mu_d^*,\Theta),
	\end{aligned}
\end{equation}
where $\mathcal L_d$ is the Lagrangian of \eqref{eq:solvesurrogate} at price $\boldsymbol\lambda_d$ and $(\boldsymbol x_d^*,\boldsymbol\nu_d^*,\boldsymbol\mu_d^*)$ is its optimal primal--dual solution, with $\boldsymbol x_d^*$ collecting $\boldsymbol y_d^*(\Theta)$ and the energy states. Appendix~\ref{subsec:appendix_gradient} gives the derivation of \eqref{eq:gradient} and the explicit expression of $\nabla_\Theta\mathcal L_d$. For fixed $\Theta$, the $D$ QPs \eqref{eq:solvesurrogate} are independent, and each term of \eqref{eq:gradient} uses only the primal--dual solution at its own price scenario, so each iteration solves them in parallel. We pass this closed-form gradient to Adam~\cite{kingma2015adam} to solve \eqref{eq:inverseoptimization}, clipping $\Theta$ to its admissible ranges after each step (Appendix~\ref{subsec:appendix_model}), as summarized in Algorithm~\ref{alg:identification}.

\begin{algorithm}[!t]
	\caption{Mean-field surrogate learning}
	\label{alg:identification}
	\begin{algorithmic}[1]
		\Statex \textbf{Input:} Device models \eqref{eq:ESS_ind}--\eqref{eq:ESS_ind_utilfunc}, price scenarios $\{\boldsymbol\lambda_d\}_{d=1}^D$, initial parameters $\Theta^{(0)}$, and iteration count $K$
		\Statex \textbf{Output:} Learned surrogate parameters $\Theta^{(K)}$
		\State Solve the device-level price-response problems and compute $\{V_{I,d}^{\mathrm{ref}}\}_{d=1}^D$ via \eqref{eq:ref_support_values}.
		\For{$k=0,\ldots,K-1$}
		\For{$d=1,\ldots,D$ in parallel}
		\State Solve the surrogate QP \eqref{eq:solvesurrogate} at $\Theta^{(k)}$ and record its optimal value $\widehat V_{\Theta^{(k)}}(\boldsymbol\lambda_d)$ and primal--dual solution.
		\EndFor
		\State Evaluate the loss in \eqref{eq:inverseoptimization}.
		\State Compute the gradient $\nabla_\Theta\mathcal J(\Theta^{(k)})$ via \eqref{eq:gradient} from the primal--dual solutions.
		\State Update $\Theta^{(k+1)}$ by an Adam step with this gradient, then clip it to the admissible ranges (Appendix~\ref{subsec:appendix_model}).
		\EndFor
		\State \Return $\Theta^{(K)}$.
	\end{algorithmic}
\end{algorithm}

%% file: V_Case.tex
This section first illustrates the theoretical results of Section~\ref{sec:MFanaly} with a toy example and then evaluates the proposed learning framework against other methods.

\subsection{Simulation Setup}\label{subsec:setup}

\begin{table}[!t]
	\centering
	\caption{Parameters of energy storage devices.}
	\small
	\renewcommand{\arraystretch}{1.12}
	\setlength{\tabcolsep}{3pt}
	\begin{tabular}{@{}ccc@{}}
		\toprule
		\textbf{Item} & \textbf{Value} & \textbf{Unit} \\
		\midrule
		$\overline p_i$ & $\operatorname{Unif}[15-10\alpha,15+10\alpha]$ & kW \\
		$\underline p_i$ & $\overline p_i\cdot\operatorname{Unif}[0.2-0.1\alpha,0.2+0.1\alpha]$ & kW \\
		$\overline e_i$ & $\overline p_i\cdot\operatorname{Unif}[2.5-1.5\alpha,2.5+1.5\alpha]$ & kWh \\
		$\underline e_i$ & $\overline e_i\cdot\operatorname{Unif}[0.2-0.1\alpha,0.2+0.1\alpha]$ & kWh \\
		$e_i^{\mathrm{ini}}$ & $\overline e_i\cdot\operatorname{Unif}[0.5-0.1\alpha,0.5+0.1\alpha]$ & kWh \\
		$\eta_i^{\mathrm C}$ & $\operatorname{Unif}[0.9-0.1\alpha,0.9+0.1\alpha]$ & / \\
		$\eta_i^{\mathrm D}$ & $\operatorname{Unif}[0.9-0.1\alpha,0.9+0.1\alpha]$ & / \\
		$q_i^{\mathrm D}$ & $\operatorname{Unif}[5-5\alpha,5+5\alpha]\times10^{-3}$ & \$/kWh \\
		$Q_i^{\mathrm D}$ & $\operatorname{Unif}[5-5\alpha,5+5\alpha]\times10^{-4}$ & \$/$(\mathrm{kW})^2$h \\
		\bottomrule
	\end{tabular}
	\label{tab:ESSParameter}
\end{table}

The parameter vector of each storage device is generated independently across devices according to Table~\ref{tab:ESSParameter}. The scalar $\alpha\in[0,1]$ controls the degree of heterogeneity across devices and is set to $\alpha=1$ unless otherwise specified. Each period is one hour ($\Delta=1$~h). In Sections~\ref{subsec:framework} and~\ref{subsec:scheduling}, the price scenarios are obtained from New York Independent System Operator (NYISO) Zone~J real-time locational marginal prices (LMPs) on randomly selected calendar days.\footnote{The price data are available at \url{https://github.com/JingguanLiu/Dataset-of-electricity-prices-for-mean-field-learning-paper}.} Appendix~\ref{subsec:appendix_model} lists the test platform and training settings.

\subsection{An Illustrative Toy Example}\label{subsec:toyexample}
\begin{figure}[!t]
	\centering
	\includegraphics[width=\linewidth]{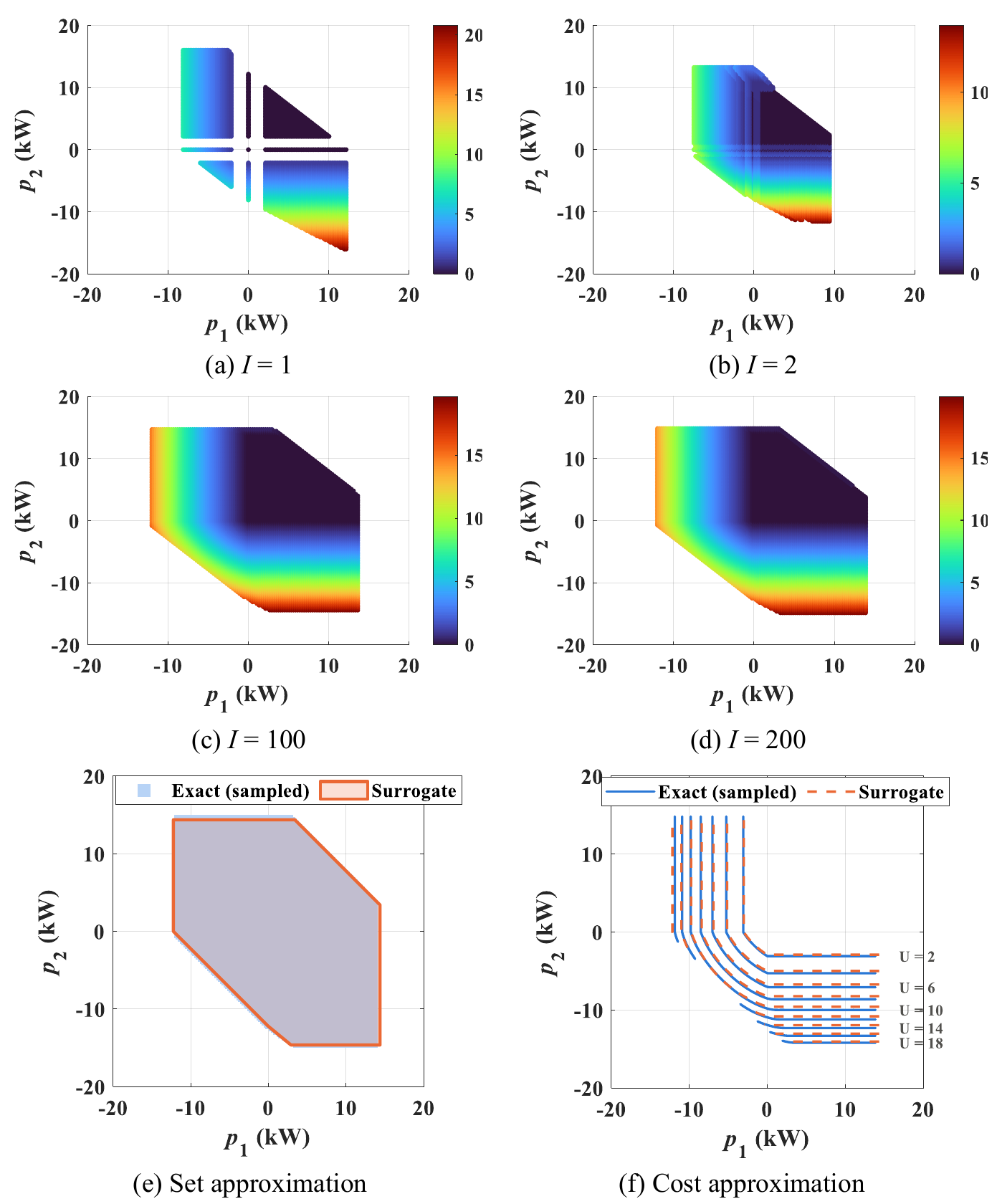}
	\caption{(a)--(d) Sampled aggregate flexibility sets $\mathcal P_I^{\mathrm M}$ for increasing population sizes, colored by the sampled disutility $U_I^{\mathrm M}$ (in cents), and (e)--(f) comparison of the learned surrogate with the $I=200$ aggregate.}
	\label{fig:toyexample_set}
\end{figure}

We use a two-period example ($T=2$) to visualize storage aggregation and surrogate approximation. Fig.~\ref{fig:toyexample_set} plots densely sampled aggregate flexibility sets over the components $p_1$ and $p_2$ of the mean net power $\boldsymbol p\in\mathcal P_I^{\mathrm M}$. The reference optimal values are computed at 50 price scenarios equally spaced on a circle in $\mathbb R^2$ of radius $50~\$/\mathrm{MWh}$, from which the training scenarios are taken.

\begin{figure}[!t]
	\centering
	\includegraphics[width=\linewidth]{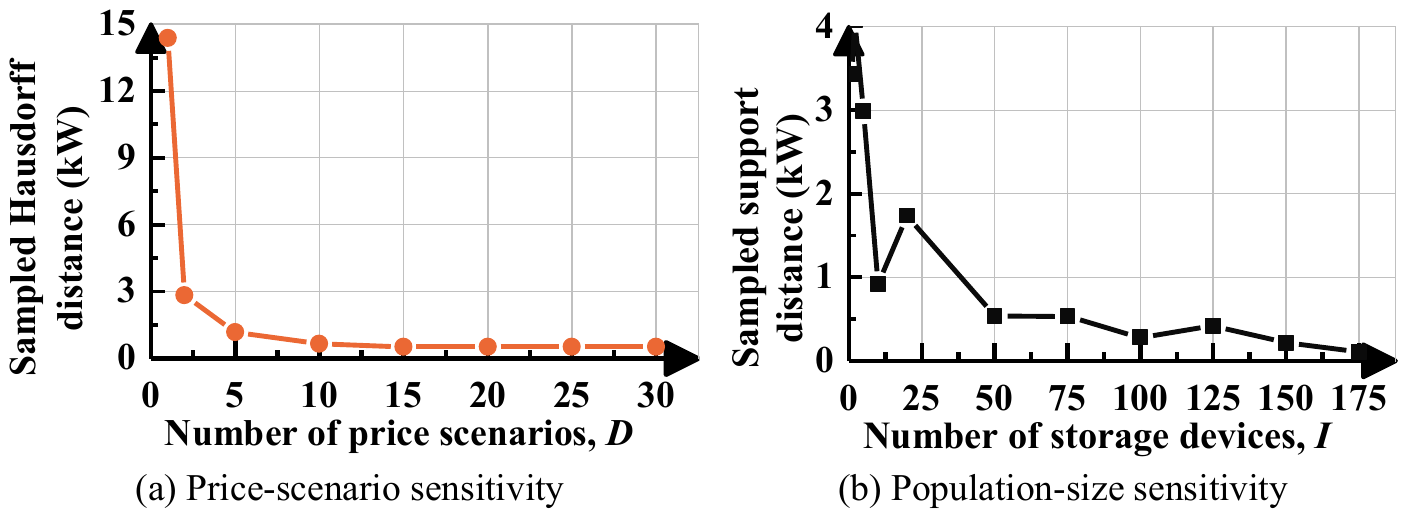}
	\caption{Sensitivity to (a) the number of training price scenarios $D$, measured by the Hausdorff distance between the flexibility sets of the learned surrogate and the sampled $I=200$ aggregate, and (b) the population size $I$, measured by the sampled support distance of random subpopulations to the $I=200$ benchmark.}
	\label{fig:toyexample_sensitive}
\end{figure}

For one realization of the device population, Fig.~\ref{fig:toyexample_set}(a)--(d) show the sampled set as the population size increases from $I=1$ to $I=200$. The $I=1$ set is disconnected and nonconvex because of the binary variables in \eqref{eq:ESS_ind}, whereas for $I=2$ most of these gaps close and the nonconvexity is less pronounced. The sets for $I=100$ and $I=200$ are nearly convex and visually similar. This pattern agrees qualitatively with the convex mean-field limit of Proposition~\ref{prop1:MFConverge}. The \textit{cooperative effect} behind it is intuitive. A single device is either idle or operates at least at its minimum power, and it cannot charge and discharge at the same time. A large fleet, however, can mix idle, charging, and discharging devices, so that its average power mimics fractional on/off states that no individual device can realize.

In Fig.~\ref{fig:toyexample_set}(e), the flexibility set of the learned surrogate nearly coincides with the sampled aggregate set for $I=200$. The disutility contours in Fig.~\ref{fig:toyexample_set}(f) also agree closely, so the surrogate captures the aggregate disutility as well.

Fig.~\ref{fig:toyexample_sensitive}(a) reports the Hausdorff distance between the flexibility sets of the learned surrogate and the sampled $I=200$ aggregate. This empirical distance measures geometric discrepancies across sampled feasible trajectories. It drops quickly and then levels off as more price scenarios are used. As discussed after Proposition~\ref{prop2:DistanceBound}, a few physical parameters determine the surrogate, so a modest number of price scenarios suffices to learn it. Fig.~\ref{fig:toyexample_sensitive}(b) takes the $I=200$ population as a finite-sample benchmark. For each $I$, a subpopulation of $I$ devices is drawn from it at random, independently across $I$, and the figure reports the sampled support distance, i.e., the largest normalized optimal-value gap $\max_d|V_{I,d}^{\mathrm{ref}}-V_{200,d}^{\mathrm{ref}}|/\|\boldsymbol\pi(\boldsymbol\lambda_d)\|$ over the 50 price scenarios on the circle. Each point comes from a single draw, so the curve fluctuates, but the distance decreases overall as the population grows, consistent with Proposition~\ref{prop3:MFConsistence}.

\subsection{Evaluation of Mean-Field Learning Framework}\label{subsec:framework}
We now consider a 24-hour horizon ($T=24$) to evaluate the proposed learning framework for $I=1000$ devices. The approximation accuracy is measured using the normalized root-mean-square error (RMSE) of the mean net-power response,
\[
\frac{\|\widehat{\boldsymbol p}_d^*(\Theta)
	-\overline{\boldsymbol p}_{I,d}^{\mathrm{ref}}\|}
{\sqrt T\,\max_t|\overline p_{I,d,t}^{\mathrm{ref}}|}\times100\%,
\]
where $\overline{\boldsymbol p}_{I,d}^{\mathrm{ref}}$ is the reference response of the fleet in \eqref{eq:ref_support_value}. For the surrogate, we also report the relative optimal-value gap $|\widehat V_\Theta(\boldsymbol\lambda_d)-V_{I,d}^{\mathrm{ref}}|/|V_{I,d}^{\mathrm{ref}}|\times100\%$, i.e., the relative counterpart of the gap minimized in \eqref{eq:inverseoptimization}.
We compare the learning framework with long short-term memory (LSTM), support vector regression (SVR), and multilayer perceptron (MLP) baselines, implemented as described in Appendix~\ref{subsec:appendix_model}. These generic regressors map the price vector to the reference net power without physical structure, whereas the surrogate is learned from the reference optimal values, so the comparison isolates the value of the surrogate's structural prior when data are scarce. All methods use the same $D=60$ training days and are tested on $300$ unseen days to assess generalization.

\begin{figure}[!t]
	\centering
	\includegraphics[width=\linewidth]{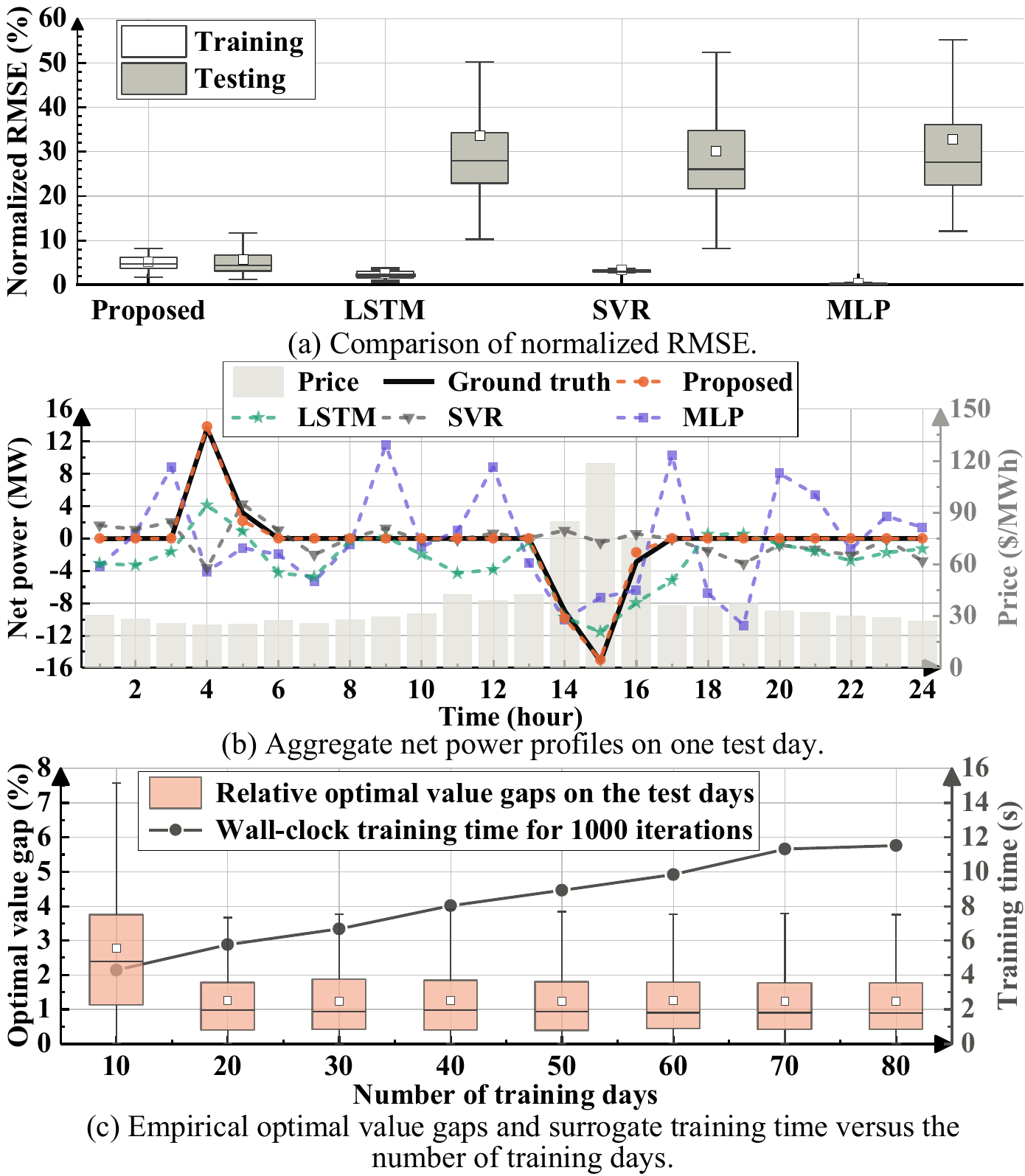}
	\caption{Learning results. In the box plots, lines mark the medians and squares mark the means. Ground truth denotes the reference aggregate net power. Power profiles are shown for the whole fleet, i.e., $I$ times the mean net power.}
	\label{fig:LearningResult}
\end{figure}

\subsubsection{Learning Results}
With $60$ days of hourly training data, the proposed framework achieves the lowest normalized RMSE on the test days (Fig.~\ref{fig:LearningResult}(a)). Its median test error is 4.4\%, close to the training median of 4.7\%, and the error stays below 10\% on 88\% of the test days. This accuracy stems from the structure of the surrogate. Guided by the convex mean-field limit, its compact parameterization \eqref{eq:VB_internal}--\eqref{eq:VB_cost} captures the essential aggregate behavior with only a few physical parameters. As a result, the estimated aggregate power profiles closely track the ground-truth aggregate trajectories, as shown in Fig.~\ref{fig:LearningResult}(b) for a test day with an afternoon price peak. In contrast, the three baselines (LSTM, SVR, MLP) fit the training days closely, with median normalized RMSEs of 0.2--3.0\%, but their median test errors are 26--28\%. Lacking the physical structure of the storage population, these models must infer the aggregate performance purely from data. They therefore generalize poorly to unseen price patterns, and their power profiles visibly deviate from the ground truth.

\subsubsection{Data and Computational Efficiency}
Fig.~\ref{fig:LearningResult}(c) reports the relative optimal-value gaps of the surrogate on the test days and its wall-clock time for surrogate learning as the number of training days increases. The median gap falls from 2.4\% with 10 days to 1.0\% with 20 days and stays at 0.9--1.0\% up to 80 days. 
As in Fig.~\ref{fig:toyexample_sensitive}(a), the gap levels off early.
Meanwhile, the training time of 1000 iterations grows only from 4.3~s to 11.5~s as the number of training days increases eightfold. This efficiency arises because each iteration only solves the small population-level QPs~\eqref{eq:solvesurrogate} in parallel, whose primal--dual solutions give the gradient~\eqref{eq:gradient} in closed form.

\subsection{Comparison with Model-Based Aggregation Method}\label{subsec:scheduling}
\begin{figure}[!t]
	\centering
	\includegraphics[width=\linewidth]{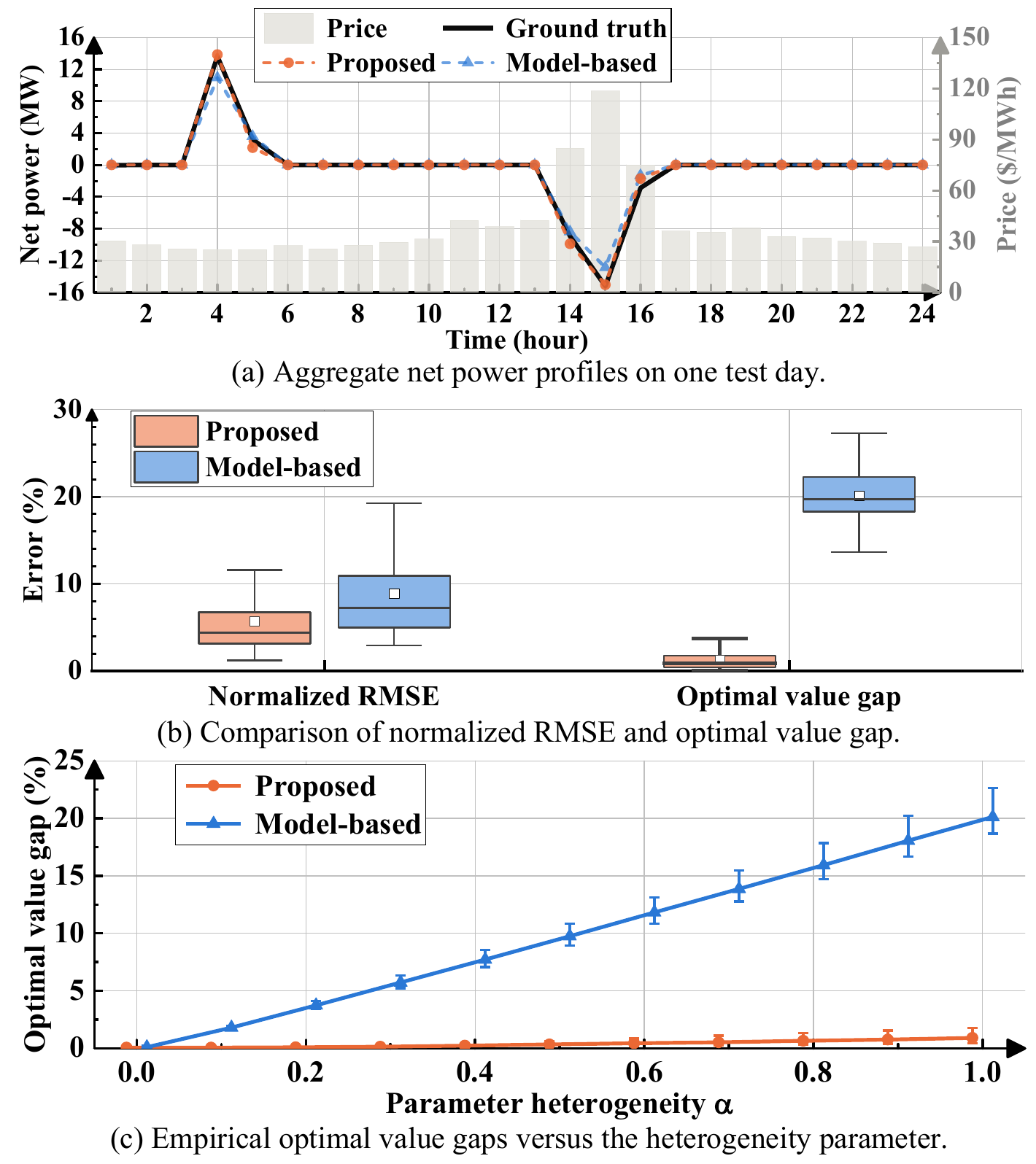}
	\caption{Proposed and model-based aggregation power profiles (whole fleet, as in Fig.~\ref{fig:LearningResult}), normalized RMSEs, and optimal-value gaps under price taking. In (c), markers show the medians over the test days at each heterogeneity level $\alpha$, offset slightly in $\alpha$ for visibility, and error bars span the quartiles.}
	\label{fig:SchedulingResult}
\end{figure}

We compare the learned surrogate $(\widehat{\mathcal P}_\Theta,\widehat U_\Theta)$ of Section~\ref{subsec:framework} with a model-based aggregation method on the same test days under price taking. For an aggregation model $(\mathcal P,U)$, the optimal value at price $\boldsymbol\lambda_d$ is $V_d(\mathcal P,U):=\min_{\boldsymbol p\in\mathcal P}\{\boldsymbol\lambda_d^\top\boldsymbol p+U(\boldsymbol p)\}$, which equals $\widehat V_\Theta(\boldsymbol\lambda_d)$ for the surrogate. For both models, Fig.~\ref{fig:SchedulingResult} reports the normalized RMSE of the optimal net power and the relative optimal-value gap of $V_d(\mathcal P,U)$, as in Section~\ref{subsec:framework}.

The baseline is a convex aggregation adapted from the general affine approximation of~\cite{al2024efficient}. Because this method requires convex device models, we relax each device's binary variables to continuous variables in $[0,1]$ for set approximation. The method then inner-approximates each relaxed device by an affine image of a common prototype virtual battery, so that the Minkowski sum of the device flexibility sets is approximated in closed form (Appendix~\ref{subsec:appendix_model}).

On the test day of Fig.~\ref{fig:LearningResult}(b), Fig.~\ref{fig:SchedulingResult}(a) shows that both aggregation methods capture the overall charging and discharging pattern of the reference, but the peaks of the baseline are lower. Over the test days, the mean daily peak power of the baseline is 14\% below that of the reference. The optimal-value approximation errors of the two methods differ substantially. In Fig.~\ref{fig:SchedulingResult}(b), the model-based baseline has a median optimal-value gap of 20.1\%, whereas the learned surrogate achieves 0.9\%, less than one-twentieth of the baseline gap, and a markedly narrower distribution.

Relaxing the binary variables lowers the devices' optimal value by only 0.25\% on average, so the baseline gap stems almost entirely from two effects of the affine approximation. First, the baseline flexibility set is too small, because the affine images of a single prototype cover only part of the flexibility sets of heterogeneous devices, which lowers the baseline peaks in Fig.~\ref{fig:SchedulingResult}(a). Second, the baseline disutility is too high. The affine maps also fix how each aggregate profile is split among the devices, and because they are chosen to cover the flexibility sets rather than to minimize cost, this split can cost more than the cheapest one. Both effects can only raise the optimal value, and the baseline indeed overestimates the reference value on every test day. Re-solving the baseline problem with the cheapest split of each profile shows that about 62\% of its average excess over the relaxed devices comes from the restricted set and 38\% from the fixed split (Appendix~\ref{subsec:appendix_model}).

We examine the effect of device heterogeneity on the optimal-value gaps by varying the heterogeneity level $\alpha$, with both models rebuilt for each value. Fig.~\ref{fig:SchedulingResult}(c) shows that the baseline gap increases with $\alpha$, while the gap of the learned surrogate remains consistently low, with a median below 1\% for every $\alpha$. With identical devices ($\alpha=0$), every affine map is the identity, so the baseline has only the relaxation error, and the median gaps are 0.11\% for the baseline and 0.06\% for the surrogate. The median gap of the baseline then grows almost linearly to 20.1\% at $\alpha=1$, as the devices differ more from the prototype.
In contrast, the proposed surrogate is learned directly from fleet-level optimal values, which reflect flexibility and disutility jointly. It is therefore less sensitive to device-level heterogeneity.

%% file: Appendix.tex
\section{Random Variables Versus Random Sets}\label{subsec:appendix_comparison}

This appendix provides an informal comparison between random variables and random sets.
Our goal is to highlight that familiar notions for random variables---such as expectation,
sample means, and the law of large numbers---admit natural counterparts for random sets,
which underpin the mean-field analysis in Section~\ref{sec:MFanaly}.
Throughout, random compact sets are assumed to be integrably bounded.

\subsection{Side-by-Side Comparison}
Table~\ref{tab:RV_RS_comparison} summarizes the key correspondences between random variables
and random sets that are directly used in our mean-field analysis.
In particular, aggregation is captured by Minkowski addition, and convergence is measured
via the Hausdorff distance.

\begin{table*}[!t]
	\centering
	\caption{Random variables versus random sets.}
	\label{tab:RV_RS_comparison}
	\renewcommand{\arraystretch}{1.2}
	\small
	\begin{tabularx}{\textwidth}{>{\raggedright\arraybackslash}p{0.24\textwidth} >{\raggedright\arraybackslash}p{0.30\textwidth} >{\raggedright\arraybackslash}X}
		\toprule
		\textbf{Concept} &
		\textbf{Random variable $\boldsymbol{x}$} &
		\textbf{Random set $\mathcal{X}$} \\
		\midrule
		Object (value space)
		&
		$\boldsymbol{x}:\Omega\rightarrow\mathbb{R}^n$, a measurable map
		taking values in $\mathbb{R}^n$
		&
		$\mathcal{X}:\Omega\rightarrow 2^{\mathbb{R}^n}$, a measurable map
		taking values in nonempty compact subsets of $\mathbb{R}^n$
		(Definition~\ref{def1:RandomSet}) \\
		\midrule
		Realization
		&
		For a fixed outcome $\omega$ in the probability space $\Omega$, $\boldsymbol{x}(\omega)\in\mathbb{R}^n$ is a
		single point
		&
		For a fixed outcome $\omega$ in the probability space $\Omega$, $\mathcal{X}(\omega)\subset\mathbb{R}^n$ is a
		single set \\
		\midrule
		Addition / aggregation
		&
		Algebraic sum:
		$\displaystyle \sum_{i=1}^I \boldsymbol{x}_i$
		&
		Minkowski sum (set addition):
		$\displaystyle \bigoplus_{i=1}^I \mathcal{X}_i$ \\
		\midrule
		Sample mean
		&
		Mean-field variable:
		$\displaystyle \overline{\boldsymbol{x}}_I
		:= \frac{1}{I}\sum_{i=1}^I \boldsymbol{x}_i$
		&
		Mean-field set:
		$\displaystyle \overline{\mathcal{X}}_I
		:= \frac{1}{I}\bigoplus_{i=1}^I \mathcal{X}_i$ \\
		\midrule
		Expectation
		&
		Expected value:
		$\displaystyle \mathbb{E}[\boldsymbol{x}]$
		&
		Aumann expectation:
		$\displaystyle \mathbb{E}[\mathcal{X}]
		:= \bigl\{ \mathbb{E}[\boldsymbol{x}] \,\bigm|\,
		\boldsymbol{x} \text{ is an integrable measurable selection of }\mathcal{X} \bigr\}$.
		The strong-law limit in Lemma~\ref{lemma1:LargeNumber} is
		$\displaystyle \mathbb{E}\bigl[\operatorname{Conv}(\mathcal{X})\bigr]$. \\
		\midrule
		Distance measure
		&
		Norm difference:
		$\bigl\|\overline{\boldsymbol{x}}_I - \mathbb{E}[\boldsymbol{x}]\bigr\|$
		&
		Hausdorff distance:
		$\displaystyle d_{\mathrm{H}}\!\left(
		\overline{\mathcal{X}}_I,\;
		\mathbb{E}\bigl[\operatorname{Conv}(\mathcal{X})\bigr]
		\right)$ \\
		\midrule
		Strong law of large numbers
		&
		$\displaystyle
		\overline{\boldsymbol{x}}_I
		\xrightarrow[I\rightarrow\infty]{\mathrm{a.s.}}
		\mathbb{E}[\boldsymbol{x}]$
		&
		$\displaystyle
		d_{\mathrm{H}}\!\left(
		\overline{\mathcal{X}}_I,\;
		\mathbb{E}\bigl[\operatorname{Conv}(\mathcal{X})\bigr]
		\right)
		\xrightarrow[I\rightarrow\infty]{\mathrm{a.s.}} 0$
		(Lemma~\ref{lemma1:LargeNumber}) \\
		\bottomrule
	\end{tabularx}
\end{table*}

\subsection{From Points to Sets}
The key distinction is that a random variable realizes a \emph{point} in $\mathbb{R}^n$,
whereas a random set realizes a \emph{compact set} in $\mathbb{R}^n$ for each outcome $\omega$.
Accordingly, the appropriate notion of expectation is the Aumann expectation, defined through
measurable selections of the random set.
Lemma~\ref{lemma1:LargeNumber} further implies that Minkowski averaging of random sets leads,
in the large-population limit, to the Aumann expectation of the convexified set
$\operatorname{Conv}(\mathcal{X})$ under Hausdorff convergence.
This is the precise sense in which random-set aggregation mirrors the classical sample-mean
convergence for random variables, while yielding a deterministic limit suitable for
population-level modeling. The example below shows how the convexification arises.

\subsection{A One-Dimensional Example}
To illustrate the analogy, consider a one-dimensional example in which a device is either idle
or operates at its rated power. Let $\xi$ be a random rated power with values in
$(0,\bar\xi]$ and define the random set
\[
\mathcal{X}(\omega):=\{0,\ \xi(\omega)\}.
\]
Each realization of $\mathcal{X}$ is thus a two-point, nonconvex set.

Now consider $I$ i.i.d.\ random sets $\mathcal{X}_i=\{0,\xi_i\}$. The mean-field set
\[
\frac{1}{I}\bigoplus_{i=1}^I \mathcal{X}_i
=\Bigl\{\frac{1}{I}\sum_{i\in\mathcal K}\xi_i\,\Bigm|\,
\mathcal K\subseteq\{1,\ldots,I\}\Bigr\}
\]
collects the average powers obtained by switching on any subset $\mathcal K$ of the devices.
It lies in the interval $[0,\frac{1}{I}\sum_{i=1}^I\xi_i]$, and switching the devices on one
at a time shows that every point of this interval is within $\bar\xi/I$ of the set. Since
$\frac{1}{I}\sum_{i=1}^I\xi_i\to\mathbb{E}[\xi]$ almost surely by the classical strong law
of large numbers,
\[
d_{\mathrm H}\!\left(\frac{1}{I}\bigoplus_{i=1}^I\mathcal{X}_i,\;[0,\mathbb{E}[\xi]]\right)
\xrightarrow[I\to\infty]{\mathrm{a.s.}}0.
\]
Every finite average is a finite set, yet the limit is the interval
$[0,\mathbb{E}[\xi]]=\mathbb{E}[\operatorname{Conv}(\mathcal{X}_1)]$, as
Lemma~\ref{lemma1:LargeNumber} predicts. Averaging many on/off decisions fills the gaps
between them, and the same mechanism makes the mean-field limit in
Proposition~\ref{prop1:MFConverge} convex.

\section{Sufficient Conditions for Assumption~\ref{assumpt1:IndSetCost}}\label{subsec:appendix_params}
Assumption~\ref{assumpt1:IndSetCost} is stated for the augmented sets. A
sufficient setting in terms of device parameters is the following. Let i.i.d.\ random
parameter vectors in a finite-dimensional space enter a common device model with finitely
many operating modes, compact decision domains, and constraint and cost functions
that are continuous in the decisions and parameters within each mode, and let every parameter
value admit a feasible decision, such as the idle schedule. The augmented
sets are then i.i.d.\ and compact-valued. They are also measurable, because the continuity and compactness assumptions give the map from parameters to augmented sets a closed graph, which implies measurability~\cite{molchanov2005theory}. If the parameters also take values in a bounded set, all feasible power profiles and costs admit common deterministic bounds,
so $\bar u$ can be chosen strictly above the common cost bound. The augmented sets are then
uniformly, hence integrably, bounded, and Assumption~\ref{assumpt1:IndSetCost} holds. For
the device model~\eqref{eq:ESS_ind}, the operating modes are the patterns of the binary
variables, the idle schedule is feasible, and the parameters in Table~\ref{tab:ESSParameter} take values in bounded ranges.

\section{Proof of Proposition~\ref{prop1:MFConverge}}\label{subsec:appendix_prop1}
\subsection{Uniformly Truncated Epigraphs}
By Assumption~\ref{assumpt1:IndSetCost}, \eqref{eq:common_upper} holds for all devices
almost surely. We fix such a realization and a finite $I$ and verify
\eqref{eq:epi_minkowski_identity}. The Minkowski average
$\mathcal S:=\frac1I\bigoplus_{i=1}^I\mathcal E_i$ is compact, as a finite Minkowski sum of
compact sets. By \eqref{eq:Epi_ESSind}, a point $(\boldsymbol p,u)$ belongs to $\mathcal S$
if and only if $\boldsymbol p=\frac1I\sum_i\boldsymbol p_i$ and $u=\frac1I\sum_iu_i$ for some
$\boldsymbol p_i\in\mathcal P_i^{\mathrm E}$ and
$u_i\in[U_i^{\mathrm E}(\boldsymbol p_i),\bar u]$. For a fixed decomposition
$\{\boldsymbol p_i\}$ of $\boldsymbol p$, the attainable averages $u$ fill the interval
$[\frac1I\sum_iU_i^{\mathrm E}(\boldsymbol p_i),\bar u]$. Hence, for
$\boldsymbol p\in\mathcal P_I^{\mathrm M}$, the vertical fiber
$\{u\mid(\boldsymbol p,u)\in\mathcal S\}$ is a union of intervals with the common upper end
$\bar u$. This fiber is closed because $\mathcal S$ is compact, so the smallest lower end is attained. It is the minimum in \eqref{eq:ESS_agg_util}, i.e., $U_I^{\mathrm M}(\boldsymbol p)$, and the fiber equals $[U_I^{\mathrm M}(\boldsymbol p),\bar u]$. For $\boldsymbol p\notin\mathcal P_I^{\mathrm M}$, the
fiber is empty. Therefore $\mathcal S=\operatorname{epi}_{\bar u}U_I^{\mathrm M}$, which is
\eqref{eq:epi_minkowski_identity}.

Lemma~\ref{lemma1:LargeNumber} applied to $\{\mathcal E_i\}$, together with
\eqref{eq:epi_minkowski_identity}, gives the Hausdorff convergence in \eqref{eq:prop1_epi}.

\subsection{Limiting Set and Disutility}
Since $\operatorname{Conv}(\mathcal E_1)$ is an integrably bounded random compact convex set,
its Aumann expectation $\mathcal E^{\mathrm L}$ is nonempty, compact, and
convex~\cite{molchanov2005theory}, and so is its projection $\mathcal P^{\mathrm L}$.

We next show that $\mathcal E^{\mathrm L}$ is vertically complete up to $\bar u$, i.e.,
$(\boldsymbol p,u)\in\mathcal E^{\mathrm L}$ implies $(\boldsymbol p,v)\in\mathcal E^{\mathrm L}$
for every $v\in[u,\bar u]$. Each set $\operatorname{epi}_{\bar u}U_I^{\mathrm M}$ has this
property and converges to $\mathcal E^{\mathrm L}$ by \eqref{eq:prop1_epi}. Fix a realization
along which this convergence holds, and let $(\boldsymbol p,u)\in\mathcal E^{\mathrm L}$ and
$v\in[u,\bar u]$. Choose $(\boldsymbol p_I,u_I)\in\operatorname{epi}_{\bar u}U_I^{\mathrm M}$
converging to $(\boldsymbol p,u)$. Then
$(\boldsymbol p_I,\max\{u_I,v\})\in\operatorname{epi}_{\bar u}U_I^{\mathrm M}$ converges to
$(\boldsymbol p,v)$, so $(\boldsymbol p,v)\in\mathcal E^{\mathrm L}$.

Hence, for each $\boldsymbol p\in\mathcal P^{\mathrm L}$, the vertical fiber
$\{u\mid(\boldsymbol p,u)\in\mathcal E^{\mathrm L}\}$ is a compact interval with upper end
$\bar u$. Defining $U^{\mathrm L}(\boldsymbol p)$ as its lower end, and
$U^{\mathrm L}(\boldsymbol p):=+\infty$ for $\boldsymbol p\notin\mathcal P^{\mathrm L}$, gives
$\mathcal E^{\mathrm L}=\operatorname{epi}_{\bar u}U^{\mathrm L}$ and
$\operatorname{dom}U^{\mathrm L}=\mathcal P^{\mathrm L}$, and this lower boundary
determines $U^{\mathrm L}$ uniquely. Convexity of $\mathcal E^{\mathrm L}$ implies convexity
of $U^{\mathrm L}$. Because $\mathcal E^{\mathrm L}$ is closed, $U^{\mathrm L}$ is lower
semicontinuous. Values outside the closed set $\mathcal P^{\mathrm L}$ are $+\infty$, so it
suffices to consider sequences in $\mathcal P^{\mathrm L}$: if $\boldsymbol p_k\to\boldsymbol p$ in $\mathcal P^{\mathrm L}$ and
$U^{\mathrm L}(\boldsymbol p_k)\to c$, then the points
$(\boldsymbol p_k,U^{\mathrm L}(\boldsymbol p_k))\in\mathcal E^{\mathrm L}$ converge to
$(\boldsymbol p,c)\in\mathcal E^{\mathrm L}$, so $U^{\mathrm L}(\boldsymbol p)\le c$. This
completes the structural part of Proposition~\ref{prop1:MFConverge}.

Projection is nonexpansive for the Hausdorff distance. Since
$\operatorname{proj}_{\boldsymbol p}\operatorname{epi}_{\bar u}
U_I^{\mathrm M}=\mathcal P_I^{\mathrm M}$, \eqref{eq:prop1_epi} yields
\[
d_{\mathrm H}(\mathcal P_I^{\mathrm M},\mathcal P^{\mathrm L})
\le d_{\mathrm H}\!\left(\operatorname{epi}_{\bar u}U_I^{\mathrm M},
\mathcal E^{\mathrm L}\right)\to0,
\]
which proves \eqref{eq:prop1_set}.

\subsection{Epi-Convergence}
Recall that $U_I^{\mathrm M}$ and $U^{\mathrm L}$ equal $+\infty$ outside
$\mathcal P_I^{\mathrm M}$ and $\mathcal P^{\mathrm L}$, respectively. Epi-convergence of
$U_I^{\mathrm M}$ to $U^{\mathrm L}$ means that
(i) $U^{\mathrm L}(\boldsymbol p)\le\liminf_{I\to\infty}U_I^{\mathrm M}(\boldsymbol p_I)$ for
every sequence $\boldsymbol p_I\to\boldsymbol p$, and (ii) every $\boldsymbol p$ admits a
sequence $\boldsymbol p_I\to\boldsymbol p$ with
$\limsup_{I\to\infty}U_I^{\mathrm M}(\boldsymbol p_I)\le U^{\mathrm L}(\boldsymbol p)$. Fix a
realization along which the Hausdorff convergence in \eqref{eq:prop1_epi} holds.

For (i), let $\boldsymbol p_I\to\boldsymbol p$ and
$\ell:=\liminf_{I\to\infty}U_I^{\mathrm M}(\boldsymbol p_I)$. If $\ell=+\infty$, there is
nothing to prove. Otherwise, along a subsequence, $U_I^{\mathrm M}(\boldsymbol p_I)\to\ell$
with $\boldsymbol p_I\in\mathcal P_I^{\mathrm M}$, so the points
$(\boldsymbol p_I,U_I^{\mathrm M}(\boldsymbol p_I))$ lie in
$\operatorname{epi}_{\bar u}U_I^{\mathrm M}$. These sets converge to the compact set
$\mathcal E^{\mathrm L}$, so the points are bounded, $\ell$ is finite, and their limit
$(\boldsymbol p,\ell)$ lies in $\mathcal E^{\mathrm L}$. Hence
$\boldsymbol p\in\mathcal P^{\mathrm L}$ and $U^{\mathrm L}(\boldsymbol p)\le\ell$.

For (ii), if $\boldsymbol p\notin\mathcal P^{\mathrm L}$, then
$U^{\mathrm L}(\boldsymbol p)=+\infty$ and $\boldsymbol p_I=\boldsymbol p$ works. If
$\boldsymbol p\in\mathcal P^{\mathrm L}$, Hausdorff convergence provides points
$(\boldsymbol p_I,u_I)\in\operatorname{epi}_{\bar u}U_I^{\mathrm M}$ that converge to
$(\boldsymbol p,U^{\mathrm L}(\boldsymbol p))\in\mathcal E^{\mathrm L}$. Then
$\boldsymbol p_I\in\mathcal P_I^{\mathrm M}$ and $U_I^{\mathrm M}(\boldsymbol p_I)\le u_I$, so
$\limsup_{I\to\infty}U_I^{\mathrm M}(\boldsymbol p_I)\le U^{\mathrm L}(\boldsymbol p)$.

Conditions (i) and (ii) prove \eqref{eq:prop1_epiconv} almost surely. \QED

\section{Proof of Proposition~\ref{prop2:DistanceBound} and Corollary~\ref{cor1:LearningBound}}\label{subsec:appendix_prop2}
By Proposition~\ref{prop1:MFConverge} and the surrogate assumptions, $\mathcal E^{\mathrm L}$
and $\widehat{\mathcal E}_\Theta$ are nonempty compact convex sets. For a compact set
$\mathcal X\subset\mathbb R^{T+1}$ and a unit vector $\boldsymbol v$, let
$m_{\mathcal X}(\boldsymbol v):=\min_{\boldsymbol z\in\mathcal X}\boldsymbol v^\top\boldsymbol z$
be its minimum support, i.e., its support function under the minimization convention. Since $\boldsymbol v(\boldsymbol\lambda)$ is
$\boldsymbol\pi(\boldsymbol\lambda)$ scaled by $1/\|\boldsymbol\pi(\boldsymbol\lambda)\|$, every
price $\boldsymbol\lambda$ satisfies
$m_{\mathcal E^{\mathrm L}}(\boldsymbol v(\boldsymbol\lambda))
=V^{\mathrm L}(\boldsymbol\lambda)/\|\boldsymbol\pi(\boldsymbol\lambda)\|$ and
$m_{\widehat{\mathcal E}_\Theta}(\boldsymbol v(\boldsymbol\lambda))
=\widehat V_\Theta(\boldsymbol\lambda)/\|\boldsymbol\pi(\boldsymbol\lambda)\|$. The proof first
expresses the Hausdorff distance through price directions and then bounds it using the price scenarios.

\subsection{Hausdorff Distance in Price Directions}
For nonempty compact convex sets, the Hausdorff distance equals the largest difference of their
minimum supports over unit directions:
\begin{equation}\label{eq:proof_hausdorff_support}
	d_{\mathrm H}(\mathcal E^{\mathrm L},\widehat{\mathcal E}_\Theta)
	=\max_{\|\boldsymbol v\|=1}
	\bigl|m_{\mathcal E^{\mathrm L}}(\boldsymbol v)-m_{\widehat{\mathcal E}_\Theta}(\boldsymbol v)\bigr|.
\end{equation}
Write a unit vector as $\boldsymbol v=(\boldsymbol r,\tau)$, where $\tau$ is the weight on the
cost coordinate. The price directions $\boldsymbol v(\boldsymbol\lambda)$ are exactly the unit
vectors with $\tau>0$, where $\boldsymbol\lambda=\boldsymbol r/\tau$, and the maximum in \eqref{eq:proof_hausdorff_support} equals the supremum over these directions. Indeed, a direction
$(\boldsymbol r,0)$ is the limit of the price directions $\boldsymbol v(s\boldsymbol r)$ as
$s\to\infty$, and minimum supports are continuous. For $\tau<0$, the objective
$\boldsymbol r^\top\boldsymbol p+\tau u$ decreases in $u$. Both sets are truncated epigraphs at
the common level $\bar u$, so each contains $(\boldsymbol p,\bar u)$ for every $\boldsymbol p$
in its projection. Both minima are therefore attained at $u=\bar u$, the term $\tau\bar u$
cancels in their difference, and what remains is
\[
\Bigl|\min_{\boldsymbol p\in\mathcal P^{\mathrm L}}\boldsymbol r^\top\boldsymbol p
-\min_{\boldsymbol p\in\widehat{\mathcal P}_\Theta}\boldsymbol r^\top\boldsymbol p\Bigr|.
\]
This is zero if $\boldsymbol r=\boldsymbol0$. Otherwise, it equals $\|\boldsymbol r\|\le1$
times the difference in the direction $(\boldsymbol r/\|\boldsymbol r\|,0)$, which has
$\tau=0$. With the identities above, \eqref{eq:proof_hausdorff_support} thus becomes
\begin{equation}\label{eq:prop2_characterization}
	d_{\mathrm H}(\mathcal E^{\mathrm L},\widehat{\mathcal E}_\Theta)
	=\sup_{\boldsymbol\lambda\in\mathbb R^T}
	\frac{\bigl|V^{\mathrm L}(\boldsymbol\lambda)-\widehat V_\Theta(\boldsymbol\lambda)\bigr|}
	{\|\boldsymbol\pi(\boldsymbol\lambda)\|}.
\end{equation}

\subsection{Gap and Coverage Errors}
For a price direction $\boldsymbol v$, the minimum $m_{\mathcal X}(\boldsymbol v)$,
$\mathcal X\in\{\mathcal E^{\mathrm L},\widehat{\mathcal E}_\Theta\}$, is attained on the graph
of the corresponding disutility, whose points have norm at most $R$. If $\boldsymbol z'$ is such
a minimizer for another price direction $\boldsymbol v'$, then
$m_{\mathcal X}(\boldsymbol v)-m_{\mathcal X}(\boldsymbol v')
\le(\boldsymbol v-\boldsymbol v')^\top\boldsymbol z'\le R\|\boldsymbol v-\boldsymbol v'\|$,
and by symmetry
$|m_{\mathcal X}(\boldsymbol v)-m_{\mathcal X}(\boldsymbol v')|\le R\|\boldsymbol v-\boldsymbol v'\|$.

Now fix $\boldsymbol\lambda$ and a price scenario $\boldsymbol\lambda_d$, and write
$\boldsymbol v=\boldsymbol v(\boldsymbol\lambda)$ and $\boldsymbol v_d=\boldsymbol v(\boldsymbol\lambda_d)$.
By the triangle inequality, the difference of the two minimum supports at $\boldsymbol v$ is at
most their difference at $\boldsymbol v_d$, which equals
$\varepsilon_d/\|\boldsymbol\pi(\boldsymbol\lambda_d)\|$, plus the change of each minimum support
from $\boldsymbol v_d$ to $\boldsymbol v$, which is at most $R\|\boldsymbol v-\boldsymbol v_d\|$.
With the identities above and $\|\boldsymbol\pi(\boldsymbol\lambda_d)\|\ge1$, this gives
\[
\begin{aligned}
	\frac{\bigl|V^{\mathrm L}(\boldsymbol\lambda)-\widehat V_\Theta(\boldsymbol\lambda)\bigr|}
	{\|\boldsymbol\pi(\boldsymbol\lambda)\|}
	&\le\frac{\varepsilon_d}{\|\boldsymbol\pi(\boldsymbol\lambda_d)\|}
	+2R\|\boldsymbol v-\boldsymbol v_d\|\\
	&\le\max_{d'}\varepsilon_{d'}+2R\|\boldsymbol v-\boldsymbol v_d\|.
\end{aligned}
\]
Minimizing over $d$ and taking the supremum over $\boldsymbol\lambda$ with
\eqref{eq:prop2_characterization} proves \eqref{eq:finite_epigraph_mse}. 
\QED

\subsection{Proof of Corollary~\ref{cor1:LearningBound}}
By the triangle inequality, every $d$ satisfies
$\varepsilon_d\le|V_{I,d}^{\mathrm{ref}}-\widehat V_\Theta(\boldsymbol\lambda_d)|
+|V_{I,d}^{\mathrm{ref}}-V^{\mathrm L}(\boldsymbol\lambda_d)|$. By
\eqref{eq:inverseoptimization}, the first term is at most
$\bigl(\sum_{d'=1}^D|V_{I,d'}^{\mathrm{ref}}-\widehat V_\Theta(\boldsymbol\lambda_{d'})|^2\bigr)^{1/2}
=\sqrt{D\,\mathcal J(\Theta)}$. Substituting into \eqref{eq:finite_epigraph_mse} gives
\eqref{eq:learning_bound}. Since $D$ is finite, \eqref{eq:support_slln} implies that the
last term in \eqref{eq:learning_bound} converges to zero almost surely. \QED

\section{Proof of Proposition~\ref{prop3:MFConsistence}}\label{subsec:appendix_prop3}
Fix $d$, write $\boldsymbol\pi_d:=\boldsymbol\pi(\boldsymbol\lambda_d)$, and let
$s_i:=\min_{\boldsymbol z\in\mathcal E_i}\boldsymbol\pi_d^\top\boldsymbol z$ denote the
optimal value of device $i$, which exists because $\mathcal E_i$ is compact.

\textit{Device-wise decomposition.}
Every point of $\frac1I\bigoplus_{i=1}^I\mathcal E_i$ has the form
$\frac1I\sum_{i=1}^I\boldsymbol z_i$ with $\boldsymbol z_i\in\mathcal E_i$. Since the objective
is linear, it is minimized over this set by minimizing each $\boldsymbol\pi_d^\top\boldsymbol z_i$
separately. Hence
$V_{I,d}^{\mathrm{ref}}=\frac1I\sum_{i=1}^Is_i$, which is \eqref{eq:ref_support_values}.

\textit{Convergence.}
The values $s_i$ are i.i.d.\ because the sets $\mathcal E_i$ are, and they are integrable
because $|s_i|\le\|\boldsymbol\pi_d\|\,\|\mathcal E_i\|$ and
$\mathbb E[\|\mathcal E_1\|]<\infty$ by Assumption~\ref{assumpt1:IndSetCost}. The strong law
of large numbers therefore gives $V_{I,d}^{\mathrm{ref}}\to\mathbb E[s_1]$ almost surely. To
identify this limit, note that a linear objective has the same minimum over $\mathcal E_1$ and
over $\operatorname{Conv}(\mathcal E_1)$, and that its minimum over an Aumann expectation
equals the expectation of its minimum over the random set~\cite{molchanov2005theory}. Hence
\[
\mathbb E[s_1]
=\mathbb E\Bigl[\min_{\boldsymbol z\in\operatorname{Conv}(\mathcal E_1)}
\boldsymbol\pi_d^\top\boldsymbol z\Bigr]
=\min_{\boldsymbol z\in\mathcal E^{\mathrm L}}\boldsymbol\pi_d^\top\boldsymbol z
=V^{\mathrm L}(\boldsymbol\lambda_d),
\]
which proves \eqref{eq:support_slln}. Because $D$ is finite, the convergence holds
simultaneously for all $d$. \QED

\section{Gradient Derivation}\label{subsec:appendix_gradient}
Fix a price $\boldsymbol\lambda_d$ and suppress it from the arguments. Write
the QP \eqref{eq:solvesurrogate} in the standard form
\begin{equation}\label{eq:kkt_standard}
	\min_{\boldsymbol x}\ f(\boldsymbol x,\Theta)
	\quad\mathrm{s.t.}\quad
	\boldsymbol h(\boldsymbol x,\Theta)=\boldsymbol0,\qquad
	\boldsymbol g(\boldsymbol x,\Theta)\le\boldsymbol0,
\end{equation}
where $\boldsymbol x$ collects $\boldsymbol p^{\mathrm C}$,
$\boldsymbol p^{\mathrm D}$, and $e_0,\ldots,e_T$, and $f$ is the objective of
\eqref{eq:solvesurrogate}. The vector $\boldsymbol h$ collects $e_0-e^{\mathrm{ini}}=0$
and the dynamics
$e_t-e_{t-1}-\Delta(\eta^{\mathrm C}p_t^{\mathrm C}-p_t^{\mathrm D}/\eta^{\mathrm D})=0$,
and $\boldsymbol g$ collects the energy and power bounds. With multipliers
$\boldsymbol\nu$ and $\boldsymbol\mu\ge\boldsymbol0$, the Lagrangian is
\[
\mathcal L(\boldsymbol x,\boldsymbol\nu,\boldsymbol\mu,\Theta)
=f(\boldsymbol x,\Theta)
+\boldsymbol\nu^\top\boldsymbol h(\boldsymbol x,\Theta)
+\boldsymbol\mu^\top\boldsymbol g(\boldsymbol x,\Theta).
\]
For fixed $\Theta$, \eqref{eq:kkt_standard} is a convex
QP with constraints affine in $\boldsymbol x$, so the KKT conditions are
necessary and sufficient for optimality.

\textit{Envelope theorem.}
Let $\widehat V(\Theta)$ denote the optimal value of \eqref{eq:kkt_standard},
i.e., $\widehat V_\Theta(\boldsymbol\lambda_d)$ in \eqref{eq:solvesurrogate}.
If \eqref{eq:kkt_standard} satisfies Slater's condition and has a unique
primal--dual solution $(\boldsymbol x^*,\boldsymbol\nu^*,\boldsymbol\mu^*)$,
then $\widehat V$ is differentiable at $\Theta$~\cite{bonnans2000perturbation} and
\begin{equation}\label{eq:envelope}
	\nabla_\Theta\widehat V(\Theta)
	=\nabla_\Theta\mathcal L(\boldsymbol x^*,\boldsymbol\nu^*,\boldsymbol\mu^*,\Theta).
\end{equation}
Intuitively, a change in $\Theta$ affects the
optimal value only through its direct effect on the objective and the
constraints at the optimum. The induced change in $\boldsymbol x^*$ has no
first-order effect because of stationarity and complementary slackness.
Applying \eqref{eq:envelope} at each price $\boldsymbol\lambda_d$ and the chain rule
to $\mathcal J(\Theta)$ in \eqref{eq:inverseoptimization} gives \eqref{eq:gradient}.

\textit{Closed form.}
Denote by $\nu_0$ the multiplier of $e_0=e^{\mathrm{ini}}$, by $\nu_t$ that of
the dynamics at period $t$, and by
$\mu_t^{\overline e},\mu_t^{\underline e},\mu_t^{\mathrm C},\mu_t^{\mathrm D}\ge0$
those of $e_t\le\overline e$, $e_t\ge\underline e$,
$p_t^{\mathrm C}\le\overline p^{\mathrm C}$, and
$p_t^{\mathrm D}\le\overline p^{\mathrm D}$. The partial derivatives of $\mathcal L$ at the optimum are
\[
\begin{aligned}
	\frac{\partial\mathcal L}{\partial q^{\mathrm D}}&{}=\textstyle\sum_{t=1}^Tp_t^{\mathrm D,*},\qquad
	\dfrac{\partial\mathcal L}{\partial Q^{\mathrm D}}=\textstyle\sum_{t=1}^T(p_t^{\mathrm D,*})^2,\\
	\frac{\partial\mathcal L}{\partial\eta^{\mathrm C}}
	&=-\Delta\textstyle\sum_{t=1}^T\nu_t p_t^{\mathrm C,*},\\
	\frac{\partial\mathcal L}{\partial\eta^{\mathrm D}}
	&=-\Delta(\eta^{\mathrm D})^{-2}\textstyle\sum_{t=1}^T\nu_t p_t^{\mathrm D,*},\\
	\frac{\partial\mathcal L}{\partial e^{\mathrm{ini}}}&=-\nu_0,\qquad
	\frac{\partial\mathcal L}{\partial\overline e}
	=-\textstyle\sum_{t=1}^T\mu_t^{\overline e},\\
	\frac{\partial\mathcal L}{\partial\underline e}
	&=\textstyle\sum_{t=1}^T\mu_t^{\underline e},\qquad
	\dfrac{\partial\mathcal L}{\partial\overline p^{\mathrm C}}
	=-\textstyle\sum_{t=1}^T\mu_t^{\mathrm C},\\
	\frac{\partial\mathcal L}{\partial\overline p^{\mathrm D}}
	&=-\textstyle\sum_{t=1}^T\mu_t^{\mathrm D}.
\end{aligned}
\]
Substituted into \eqref{eq:gradient}, these expressions give the gradient from
the optimal primal--dual solutions of \eqref{eq:solvesurrogate} alone. Algorithm~\ref{alg:identification} uses this gradient to solve \eqref{eq:inverseoptimization}.

\section{Implementation Details}\label{subsec:appendix_model}
All numerical experiments are conducted on a laptop with an Intel Core i9-14900HX CPU (2.20~GHz, 24 cores) and 32~GB of RAM. Mixed-integer reference problems are solved using \texttt{Gurobi}. At each iteration, the optimal primal--dual solutions of the surrogate QPs \eqref{eq:solvesurrogate} are held fixed, and the Lagrangians $\mathcal L_d$ are evaluated in \texttt{PyTorch} as functions of $\Theta$. Back-propagating the loss through these Lagrangians returns the closed-form gradient \eqref{eq:gradient}, which is passed to Adam, so no differentiation through the QP solver is needed. In Section~\ref{subsec:framework}, the QPs of each iteration are solved in parallel by 24 worker processes.  We use Adam with a learning rate of 0.01 for 100 iterations (1000 iterations in Sections~\ref{subsec:framework} and~\ref{subsec:scheduling}). Because shifting $e^{\mathrm{ini}}$, $\underline e$, and $\overline e$ by a common constant leaves $\mathcal Y_\Theta$ in \eqref{eq:VB_internal} unchanged, the implementation fixes $e^{\mathrm{ini}}=0$. After each Adam step, $\eta^{\mathrm C}$ and $\eta^{\mathrm D}$ are clipped to $[0.7,1]$, $q^{\mathrm D}$ is kept nonnegative, and $\overline p^{\mathrm C}$, $\overline p^{\mathrm D}$, $\overline e$, $-\underline e$, and $Q^{\mathrm D}$ are kept positive.

\subsection{Predictive Baselines}
Baseline models are configured as follows:
\begin{itemize}
	\item \textbf{LSTM:} A two-layer bidirectional LSTM with 32 units per direction reads the 24 hourly prices as a sequence, and a fully connected layer maps its output at each hour to the aggregate power of that hour.
	\item \textbf{SVR:} One model with a radial basis function kernel ($C=100$, $\epsilon=0.1$) is trained for each hour. It receives the 24-hour price vector and predicts the aggregate power of that hour.
	\item \textbf{MLP:} Two hidden layers of 64 rectified linear units map the 24-hour price vector to the aggregate power profile.
\end{itemize}
Inputs and outputs are standardized per hour with the statistics of the training days. The LSTM and MLP minimize the mean squared error with Adam (learning rate $10^{-3}$, weight decay $10^{-4}$) over 2000 full-batch epochs.

\subsection{Model-Based Baseline}\label{subsec:appendix_mb}
The baseline in Section~\ref{subsec:scheduling} adapts the affine inner approximation of~\cite{al2024efficient} to storage devices. Relaxing $\delta_{i,t}^{\mathrm C},\delta_{i,t}^{\mathrm D}\in\{0,1\}$ in \eqref{eq:ESS_ind_d} to $[0,1]$ removes the minimum power limit and turns charge--discharge exclusivity into the shared limit $p_{i,t}^{\mathrm C}+p_{i,t}^{\mathrm D}\le\overline p_i$. With $e_{i,t}$ expressed through \eqref{eq:ESS_ind_a}, the relaxed trajectories of device~$i$ form a polytope $\widetilde{\mathcal Y}_i:=\{\boldsymbol y\in\mathbb R^{2T}\mid\boldsymbol A_i\boldsymbol y\le\boldsymbol b_i\}$ that contains $\mathcal Y_i^{\mathrm E}$. The prototype $\mathcal Y_0:=\{\boldsymbol y\in\mathbb R^{2T}\mid\boldsymbol A_0\boldsymbol y\le\boldsymbol b_0\}$ averages the device constraints, with $\boldsymbol A_0:=\frac1I\sum_{i=1}^I\boldsymbol A_i$ and $\boldsymbol b_0:=\frac1I\sum_{i=1}^I\boldsymbol b_i$. 

Each relaxed set is inner-approximated by an affine image of the prototype, $\boldsymbol\gamma_i+\boldsymbol\Gamma_i\mathcal Y_0\subseteq\widetilde{\mathcal Y}_i$, with $\boldsymbol\gamma_i\in\mathbb R^{2T}$ and $\boldsymbol\Gamma_i\in\mathbb R^{2T\times2T}$. This inclusion holds if and only if a nonnegative matrix $\boldsymbol\Lambda_i$ satisfies $\boldsymbol\Lambda_i\boldsymbol A_0=\boldsymbol A_i\boldsymbol\Gamma_i$ and $\boldsymbol\Lambda_i\boldsymbol b_0\le\boldsymbol b_i-\boldsymbol A_i\boldsymbol\gamma_i$~\cite{al2024efficient}. For each device, we maximize $\operatorname{tr}(\boldsymbol\Gamma_i)$, which linearizes the volume of the image around $\boldsymbol\Gamma_i=\boldsymbol I$, subject to these linear conditions. The resulting $I$ linear programs are independent and are solved once. 

The affine maps also fix the disaggregation. Each prototype trajectory $\boldsymbol y_0\in\mathcal Y_0$ assigns device~$i$ the trajectory $\boldsymbol y_i=\boldsymbol\gamma_i+\boldsymbol\Gamma_i\boldsymbol y_0\in\widetilde{\mathcal Y}_i$, and the average of these trajectories is $\bar{\boldsymbol\gamma}+\bar{\boldsymbol\Gamma}\boldsymbol y_0$, where $\bar{\boldsymbol\gamma}:=\frac1I\sum_{i=1}^I\boldsymbol\gamma_i$ and $\bar{\boldsymbol\Gamma}:=\frac1I\sum_{i=1}^I\boldsymbol\Gamma_i$. The baseline flexibility set $\mathcal P^{\mathrm{MB}}$ is the net-power projection of $\bar{\boldsymbol\gamma}+\bar{\boldsymbol\Gamma}\mathcal Y_0$, as in \eqref{eq:VB_set}. Since every $\boldsymbol y_i$ is feasible for the relaxed device, $\mathcal P^{\mathrm{MB}}$ is an inner approximation of the aggregate of the relaxed devices. The baseline disutility is the average device cost of the disaggregated trajectories, minimized over the prototype trajectories that realize $\boldsymbol p$:
\[
U^{\mathrm{MB}}(\boldsymbol p):=\min_{\substack{\boldsymbol y_0\in\mathcal Y_0,\ \boldsymbol y=\bar{\boldsymbol\gamma}+\bar{\boldsymbol\Gamma}\boldsymbol y_0\\
		\boldsymbol p^{\mathrm C}-\boldsymbol p^{\mathrm D}=\boldsymbol p}}
\frac1I\sum_{i=1}^Ic_i(\boldsymbol\gamma_i+\boldsymbol\Gamma_i\boldsymbol y_0),
\]
where $c_i(\boldsymbol y_i):=\sum_{t\in\mathcal T}\bigl(q_i^{\mathrm D}p_{i,t}^{\mathrm D}+Q_i^{\mathrm D}(p_{i,t}^{\mathrm D})^2\bigr)$ is the device cost in \eqref{eq:ESS_ind_utilfunc}. The price-response problem that defines $V_d(\mathcal P^{\mathrm{MB}},U^{\mathrm{MB}})$ is therefore a convex QP in $\boldsymbol y_0$. Its coefficients are averaged over the devices once, so its size does not grow with $I$.

To separate the two effects discussed in Section~\ref{subsec:scheduling}, we compare three optimal values on each test day: (i) the value $V_d^{\mathrm R}$ of the relaxed devices, (ii) the value $V_d^{\mathrm R}(\mathcal P^{\mathrm{MB}})$ of the same devices with their mean net power restricted to $\mathcal P^{\mathrm{MB}}$, which splits each baseline profile among the devices at minimum cost, and (iii) the baseline value $V_d(\mathcal P^{\mathrm{MB}},U^{\mathrm{MB}})$. Restricting the mean net power can only raise the optimal value, and fixing the split can only raise it further, so $V_d^{\mathrm R}\le V_d^{\mathrm R}(\mathcal P^{\mathrm{MB}})\le V_d(\mathcal P^{\mathrm{MB}},U^{\mathrm{MB}})$. Relative to $|V_{I,d}^{\mathrm{ref}}|$, the first difference, caused by the restricted set, averages 13.0\% over the test days, and the second, caused by the fixed split, averages 7.8\%.

\section{Stress Test with Correlated Devices}\label{subsec:appendix_noniid}
\begin{figure}[!t]
	\centering
	\includegraphics[width=\linewidth]{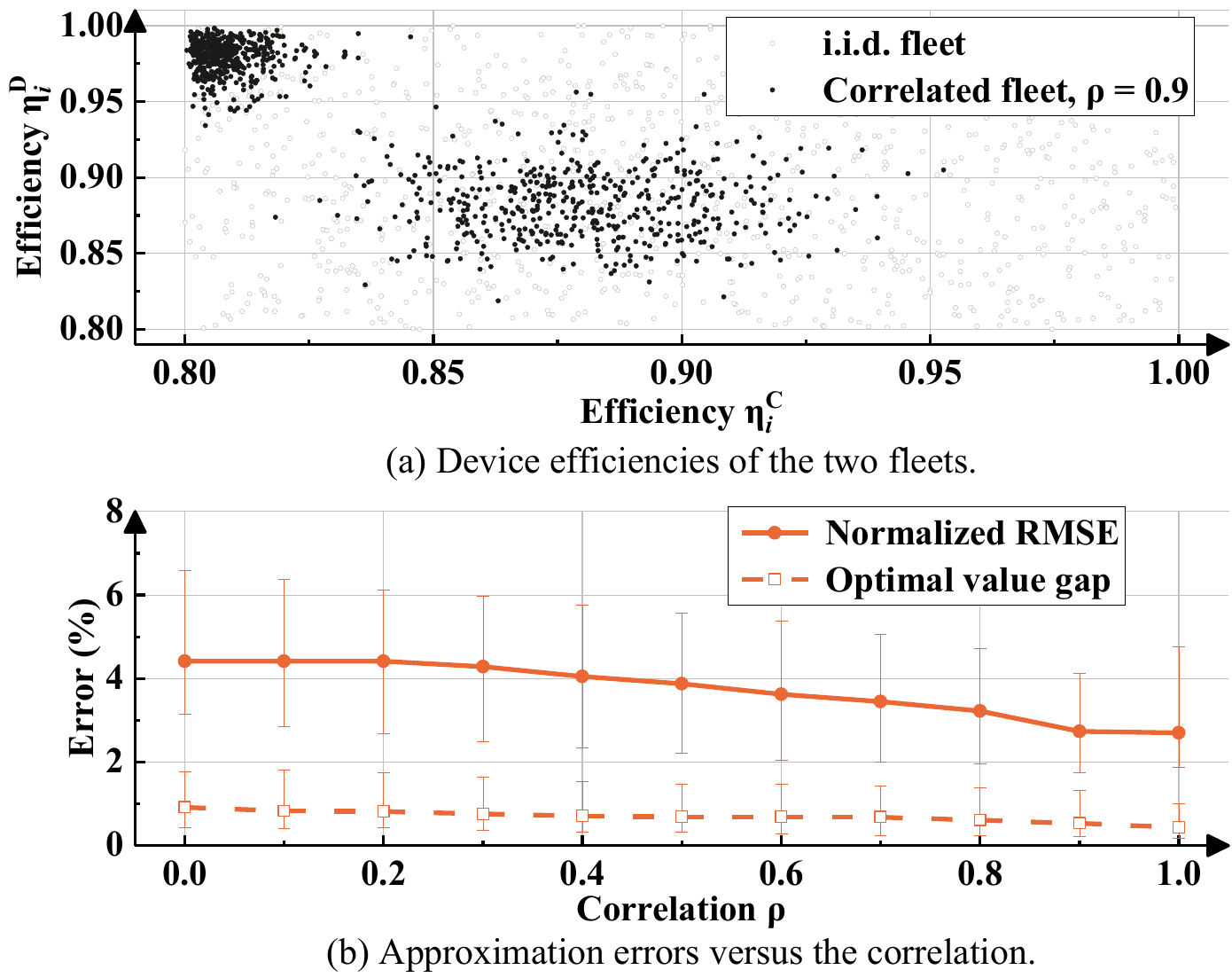}
	\caption{Stress test with correlated devices. (a) Charging ($\eta_i^{\mathrm C}$) and discharging ($\eta_i^{\mathrm D}$) efficiencies of the i.i.d.\ fleet and of the correlated fleet with $\rho=0.9$. (b) Normalized RMSE and optimal-value gap on the test days versus $\rho$. Markers show the medians, and error bars span the quartiles.}
	\label{fig:noniid}
\end{figure}

Assumption~\ref{assumpt1:IndSetCost} fails when devices share random factors, as under strong spatial or demographic clustering (Remark~\ref{rem:AssumptionScope}). To test the learning method in this case, we divide the 1000 devices of Section~\ref{subsec:framework} into two clusters of 500, such as two regions served by different installers, and correlate the devices within each cluster. Each uniform distribution in Table~\ref{tab:ESSParameter} ($\alpha=1$) is sampled as $a+(b-a)w_i$, where $[a,b]$ is its range and the standard uniform variables $w_i$ are independent across devices and parameters. For device~$i$ in cluster~$k$, we replace $w_i$ by
\[
\tilde w_i:=\Phi\bigl(\sqrt{\rho}\,\zeta_k+\sqrt{1-\rho}\,\Phi^{-1}(w_i)\bigr),
\]
where $\Phi$ is the standard normal distribution function, $\rho\in[0,1]$, and the cluster factors $\zeta_k$ are standard normal, drawn independently for each cluster and parameter. In this one-factor Gaussian copula, $\Phi^{-1}(w_i)$ and $\zeta_k$ are independent standard normal variables, so their weighted sum is standard normal and $\tilde w_i$ is again standard uniform. Each device thus keeps its marginal distribution, whereas two devices in the same cluster share $\zeta_k$, and their weighted sums have correlation~$\rho$. Setting $\rho=0$ recovers the i.i.d.\ fleet of Section~\ref{subsec:framework}, and $\rho=1$ makes the devices of a cluster identical. Given the factors, the devices of each cluster are i.i.d., so by Lemma~\ref{lemma1:LargeNumber} the mean-field aggregate converges, as both clusters grow, to the average of two cluster limits. For $\rho>0$, these limits depend on the realized factors, so the mean-field limit is random. The prices, training and test days, and learning settings are those of Section~\ref{subsec:framework}.

Fig.~\ref{fig:noniid} shows one draw of the factors. In Fig.~\ref{fig:noniid}(a), the charging and discharging efficiencies of the correlated fleet with $\rho=0.9$ concentrate in two clusters, whereas those of the i.i.d.\ fleet spread uniformly. Fig.~\ref{fig:noniid}(b) shows that the errors of the learned surrogate decrease as $\rho$ grows. From $\rho=0$ to 1, the median optimal-value gap falls from 0.9\% to 0.4\%, and the median normalized RMSE falls from 4.4\% to 2.7\%. The errors fall because correlation makes the devices of a cluster alike, and a single virtual battery represents similar devices more accurately. The diversity of the fleet, measured by the standard deviation of $\tilde w_i$ across devices and averaged over the parameters, falls from $1/\sqrt{12}\approx0.29$ for the i.i.d.\ fleet to 0.12 at $\rho=1$, where each parameter takes only two values. Section~\ref{subsec:scheduling} shows the same effect of diversity. There, the median gap of the surrogate falls from 0.9\% for $\alpha=1$ to 0.06\% for identical devices ($\alpha=0$).

Two other draws of the factors, at $\rho=0.5$, 0.9, and 1, confirm that the surrogate remains accurate for correlated devices. The median normalized RMSE is 2.7--3.4\%, below the 4.4\% of the i.i.d.\ fleet, and the median gap is at most 1.4\%, compared with 0.9\% for the i.i.d.\ fleet. The 1.4\% gap occurs at $\rho=1$ for the only draw whose two clusters differ markedly in storage duration $\overline e_i/\overline p_i$, namely 1.9~h versus 1.4~h, whereas they differ by only 0.05~h in the other two of the three draws. Charged at full power, such clusters fill up at different times, which a single virtual battery cannot reproduce exactly. Giving both clusters the same duration, with the same total power and energy, lowers this gap to 0.9\%. The rise is thus caused by the difference between the two clusters rather than by the correlation. Indeed, independent devices drawn from an equal mixture of the same two types, which satisfy Assumption~\ref{assumpt1:IndSetCost} (Remark~\ref{rem:AssumptionScope}), would form essentially the same fleet.

In summary, the learned surrogate remains accurate for correlated devices. Its accuracy is governed by how much the devices differ, whether or not they are independent. Consistent with Remark~\ref{rem1:LearningScope}, independence is needed only for the mean-field limit.